\documentclass[aps,prx,preprint,onecolumn,citeautoscript,footinbib,eqsecnum]{revtex4-1}  
\usepackage{amsmath,amssymb,bm,bbm} 
\usepackage{graphicx}  
\usepackage{float}
\usepackage[caption = false]{subfig}
\usepackage{color} 
\usepackage[dvipsnames]{xcolor}
\usepackage[papersize={8.5in,11in}]{geometry}
\usepackage[colorlinks=true]{hyperref}  
\usepackage{ulem}
\usepackage[mathscr]{euscript}
\usepackage[section]{placeins}
\usepackage{comment}
\hypersetup{
    bookmarks=true,         
    unicode=false,          
    pdftoolbar=true,        
    pdfmenubar=true,        
    pdffitwindow=false,     
    pdfstartview={FitH},    
    pdfsubject={},   
    pdfcreator={},   
    pdfproducer={}, 
    pdfkeywords={} {} {}, 
    pdfnewwindow=true,      
    colorlinks=true,       
    linkcolor=magenta, 
    citecolor=blue,        
    filecolor=magenta,      
    urlcolor=blue           
} 

\usepackage{amsfonts}
\usepackage{upgreek}
\usepackage{slashed}
\usepackage{latexsym}
\usepackage[usestackEOL]{stackengine}
\newcommand{\beq}{\begin{equation}}
\newcommand{\eeq}{\end{equation}}
\def\bea{\begin{eqnarray}}
\def\eea{\end{eqnarray}}
\newcommand{\nn}{\nonumber \\}

\begin{document}

\preprint{\href{https://arxiv.org/abs/2103.05009}{arXiv:2103.05009}}

\title{Small to large Fermi surface transition\\ in a single band model,\\ using randomly coupled ancillas}

\author{Alexander Nikolaenko}
\affiliation{Department of Physics, Harvard University, Cambridge MA 02138, USA}

\author{Maria Tikhanovskaya}
\affiliation{Department of Physics, Harvard University, Cambridge MA 02138, USA}

\author{Subir Sachdev}
\affiliation{Department of Physics, Harvard University, Cambridge MA 02138, USA}

\author{Ya-Hui Zhang}
\affiliation{Department of Physics, Harvard University, Cambridge MA 02138, USA}

\date{\today\\
\vspace{0.4in}}

\begin{abstract}
We describe a solvable model of a quantum transition in a single band model involving a change in the size of the electron Fermi surface without any symmetry breaking. 
In a model with electron density $1-p$, we find a `large' Fermi surface state with the conventional Luttinger volume $1-p$ of electrons for $p>p_c$, and a first order transition to a `small' Fermi surface state with a non-Luttinger
volume $p$ of holes for $p<p_c$. As required by extended Luttinger theorems, the small Fermi surface state also has fractionalized spinon excitations. The model has electrons with strong local interactions in a single band; after a canonical transformation, the interactions are transferred to a coupling  to two layers of ancilla qubits, as proposed by Zhang and Sachdev (Phys. Rev. Research {\bf 2}, 023172 (2020)). 
Solvability is achieved by employing random exchange interactions within the ancilla layers, and taking the large $M$ limit with SU($M$) spin symmetry, as in the Sachdev-Ye-Kitaev models. The local electron spectral function of the small Fermi surface phase displays a particle-hole asymmetric pseudogap, and maps onto the spectral function of a lightly doped Kondo insulator of a Kondo-Heisenberg lattice model. We discuss connections to the physics of the hole-doped cuprates: the asymmetric pseudogap observed in STM, and the sudden change from incoherent to coherent anti-nodal spectra observed recently in photoemission. A holographic analogy to wormhole transitions between multiple black holes is briefly noted.
\end{abstract}

\maketitle
\tableofcontents

\section{Introduction}

A number of recent  experiments \cite{CPLT18,Shen19,Ramshaw20} have highlighted the role of optimal doping in the hole-doped cuprate superconductors, where there is a rapid change in the size of the underlying Fermi surface. At hole doping $p$ away from half-filling, a conventional, Fermi-liquid-like, large Fermi surface corresponding to an electron density of $1-p$ is obtained for $p> p_c$, where $p_c$ is critical optimal doping. In the pseudogap regime for $p<p_c$, a small Fermi surface corresponding to a hole density $p$ is observed.
We shall take the point-of-view here that this change in Fermi surface size is the primary physics driving the transition. Symmetry breaking is often observed at low temperatures in the small Fermi surface regime, but we shall view this here as a secondary phenomenon to be understood in a more refined treatment.

Large-to-small Fermi surface transformations without symmetry breaking, between phases that obey and violate the conventional Luttinger theorem, have been studied a great deal \cite{Burdin_2002,Senthil_2003,SVS04,Paramekanti_2004,Coleman_2005,Norman07,Norman08,Norman13,Si2010,Si2014,PaschenSi21,Chowdhury_2018,PatelAltman20}  in the context of {\it two\/} band Kondo lattice models. 
Such models have a band of localized spins coupled to a second band of mobile electrons. In the large Fermi surface phase (FL), the localized spins are Kondo screened by the conduction electrons, and so the Fermi surface size corresponds to the combined density of the mobile electrons and the localized spins \cite{MO00}. In the small Fermi surface phase, the `fractionalized Fermi liquid' (FL*), the spins form a decoupled spin liquid with fractionalized spinon excitations, and the Fermi surface size corresponds only to the density of mobile electrons. 
The conventional Luttinger relation for the Fermi surface size is obeyed only in the large Fermi surface phase. On the other hand, in the small Fermi surface phase, the fractionalized excitations and emergent gauge fields accompanying the small Fermi surface allow this phase to satisfy a generalized Luttinger relation \cite{Senthil_2003,SVS04,Paramekanti_2004,Else2020}. (We note that in studies of `Kondo breakdown' critical points \cite{Sengupta00,SiNature,QSi19,QSi20}, the small Fermi surface phase has broken symmetry and obeys the conventional Luttinger theorem, and so this phase is not required to have fractionalized excitations.)
Recent experiments on CePdAl \cite{Sun19} and CeCoIn$_5$ \cite{Analytis20} have presented significant evidence for such a small-to-large Fermi surface transition without any symmetry breaking. 

Large-to-small Fermi surface transformations also appear in various dynamic mean-field theory treatments of multi-band models, where they are often referred to as `orbitally-selective Mott transitions' \cite{Anisimov_2002,Medici_05,Becca_2016,Si_2017,Si_2018,Anisimov_2021,Si_2021}. But these treatments do not account for the fractionalized excitations that are required to appear along with the small Fermi surface to account for any violation of the Luttinger value for the Fermi surface size.

For the cuprates, the observations appear to require a small-to-large Fermi surface transition in a {\it single} band model. Models of small Fermi surfaces of electrons obtained by the non-perturbative binding spinons and holons excitations of a doped spin liquid have been proposed \cite{SS18,WenLee96,YRZ06,TsvelikRice19,Kaul08,YQSS10,EGMSS11,Mei12,Punk_2012,Punk_2015,SCWFGS17,Punk17,Punk19,Punk_2020,SS19}, but none provide a fully self-consistent method for computing the Fermi surface in both the small and large Fermi surface states. Unlike the Kondo lattice model, there is no natural criterion for choosing between the electrons which form local moments and fractionalize, and those which are mobile, and so this leads to significant technical difficulties in obtaining the small Fermi surface FL* state. 

Recent work \cite{Yahui20a,Yahui20b} has shown that many of these difficulties are overcome in an `ancilla qubit' approach, which can describe both the small and large Fermi surface states of a single band model. This approach begins with a single band Hubbard model of electrons, $C$. We then perform an analog of a Hubbard-Stratonovich transformation on the Hubbard interaction, by `integrating in' a pair of $S=1/2$ ancilla spins on each site, coupled to each other by a large antiferromagnetic exchange coupling $J_\perp$ (see Fig.~\ref{fig:layers}). As described in Appendix~\ref{app:hubbard}, upon eliminating the ancillas by a canonical transformation, in a $1/J_\perp$ expansion which locks the ancillas into rung spin singlets, 
we recover the original single band Hubbard model of $C$ electrons. But we choose instead to keep the ancilla degrees of freedom `alive' at intermediate stages, and work in the canonically equivalent model of {\it free\/} electrons coupled to the ancillas: this gives us the flexibility needed to obtain a large $M$ saddle point which describes the pseudogap phase.
Fluctuations of a SU(2)$_S$ gauge field beyond the saddle-point are needed to project the ancillas into a rung-singlet subspace \cite{Yahui20a,Yahui20b}, and this ensures that the final theory is expressed only in terms of the single band degrees of freedom of the physical $C$ electrons. The theory of the SU(2)$_S$ gauge fluctuations shows that the small Fermi surface state is stable to the projection to the physical degrees of freedom: this is because the SU(2)$_S$ gauge fluctuations are higgsed in the FL* phase.

In the present paper, we will combine the ancilla qubit method for a single band model, with the method employed by Burdin {\it et al.\/} \cite{Burdin_2002} for the Kondo lattice model. We will couple two ancilla layers to a physical single band model, and use a Sachdev-Ye-Kitaev \cite{sy,Kitaev2015,SS15,GKST} (SYK) description of the spin liquid states on the ancillas; this is achieved by including a random exchange interaction of mean-square strength $J$ {\it within} each ancilla layer. We show below that this leads to a tractable 
description of the phases on both sides of a first order Fermi volume changing transition, while also providing a self-consistent description of the incoherent and fractionalized excitations. 

We begin by recalling the ancilla qubit method and its description of the phases of the single band model \cite{Yahui20a,Yahui20b}: see Fig.~\ref{fig:layers}.
\begin{figure}
\includegraphics[width=5in]{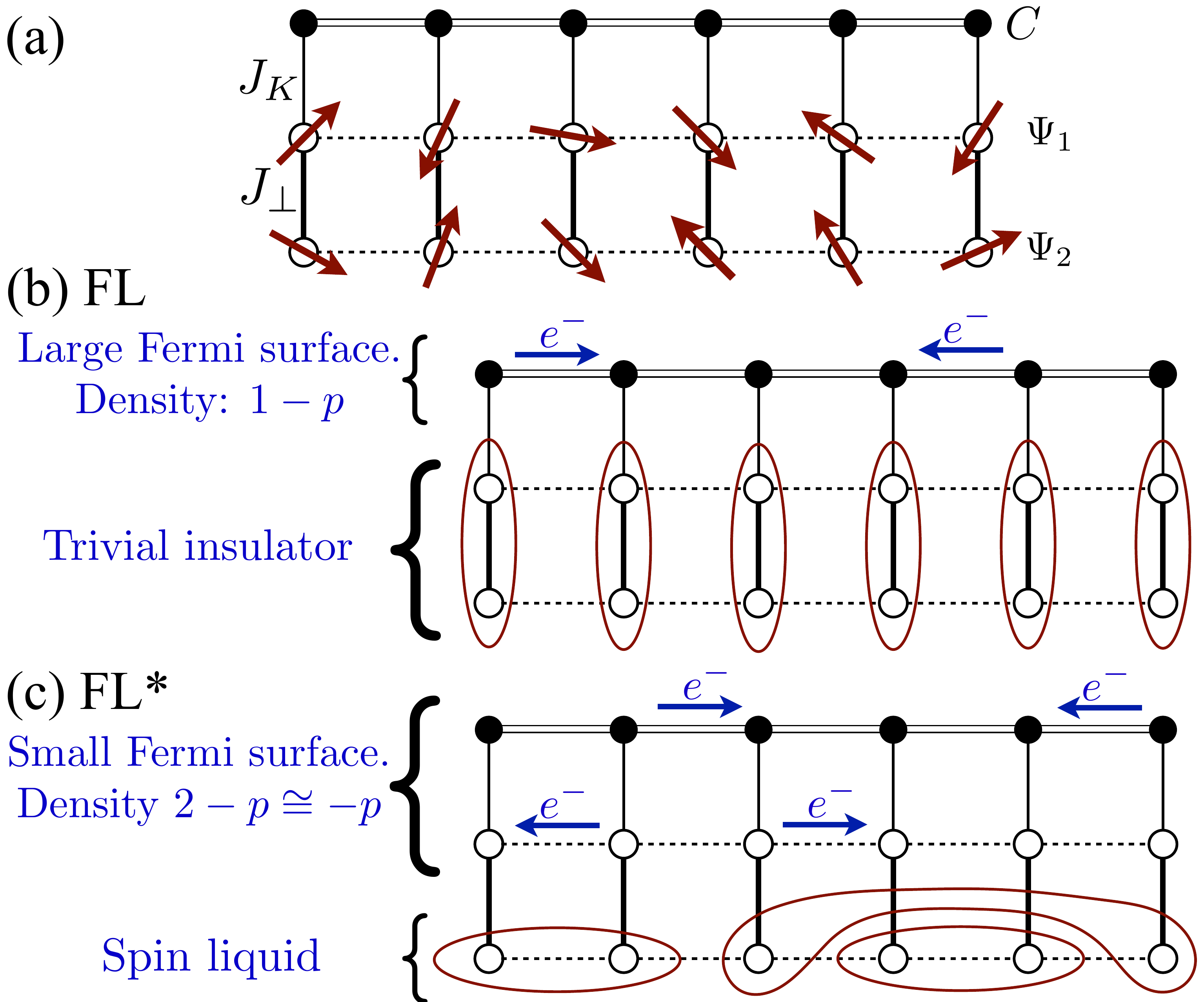}
\caption{(a) The top layer is the physical layer of electrons $C$ in a single band model coupled to two `hidden' layers of ancilla qubits (spin-1/2 spins) realized by fermions $\Psi_1$ and $\Psi_2$. The antiferromagnetic exchange couplings $J_K$ and $J_\perp$ are non-random, while the dashed lines represent random exchange interactions of mean-square strength $J$ between the $\Psi_1$ spins and between the $\Psi_2$ spins. (b) In the large Fermi surface FL phase, the ancilla spins lock into rung singlets, while the $C$ electrons are largely decoupled from the ancilla and form a conventional Fermi liquid of electron density $1-p$. (c) In the small Fermi surface FL* phase, the $\Psi_1$ ancilla spins are Kondo screened by the $C$ electrons to form a Fermi surface with density $2-p$ electrons. This is equivalent to a small hole-like Fermi surface of size $p$, as observed in the cuprates at low doping. The $\Psi_2$ spins are largely decoupled from the top two layers in the FL* phase, and form a gapless spin liquid with fractionalization, whose presence is required by the generalized Luttinger theorem. }
\label{fig:layers}
\end{figure}
The top physical layer of electrons, $C$, of density $1-p$ is coupled to 2 layers of ancilla qubits. The ancilla qubits are realized by fermions $\Psi_{1,2}$ using the usual Schwinger construction, with the constraint $\sum_{\alpha} \Psi_{i;a;\alpha}^\dagger \Psi_{i;a; \alpha} = 1$ satisfied on each lattice site $i$ ($a=1,2$ is a layer index, and $\alpha = \uparrow, \downarrow$ is a spin index). It is important that we add {\it two} layers of ancilla qubits, because only then are the added layers free of all anomalies \cite{Else2020}, and are allowed to form a trivial insulator. Some previous discussions of the FL* phase \cite{YQSS10,EGMSS11} were obtained by adding a single band near half-filling: this gives a suitable description of the electron spectral function in the FL* phase, but misses the FL* spectrum of spin excitations associated with the second ancilla layer, and cannot obtain a FL phase.

In the large Fermi surface FL phase, we assume that the non-random and antiferromagnetic coupling $J_\perp$ dominates, and so the ancilla are locked into rung singlets, and can be safely ignored in the low energy theory: then the $C$ electrons form a conventional Fermi liquid phase, and we obtain a Fermi surface corresponding to electron density $1-p$, or hole density $1+p$.

\begin{figure}
\includegraphics[width=4.3in]{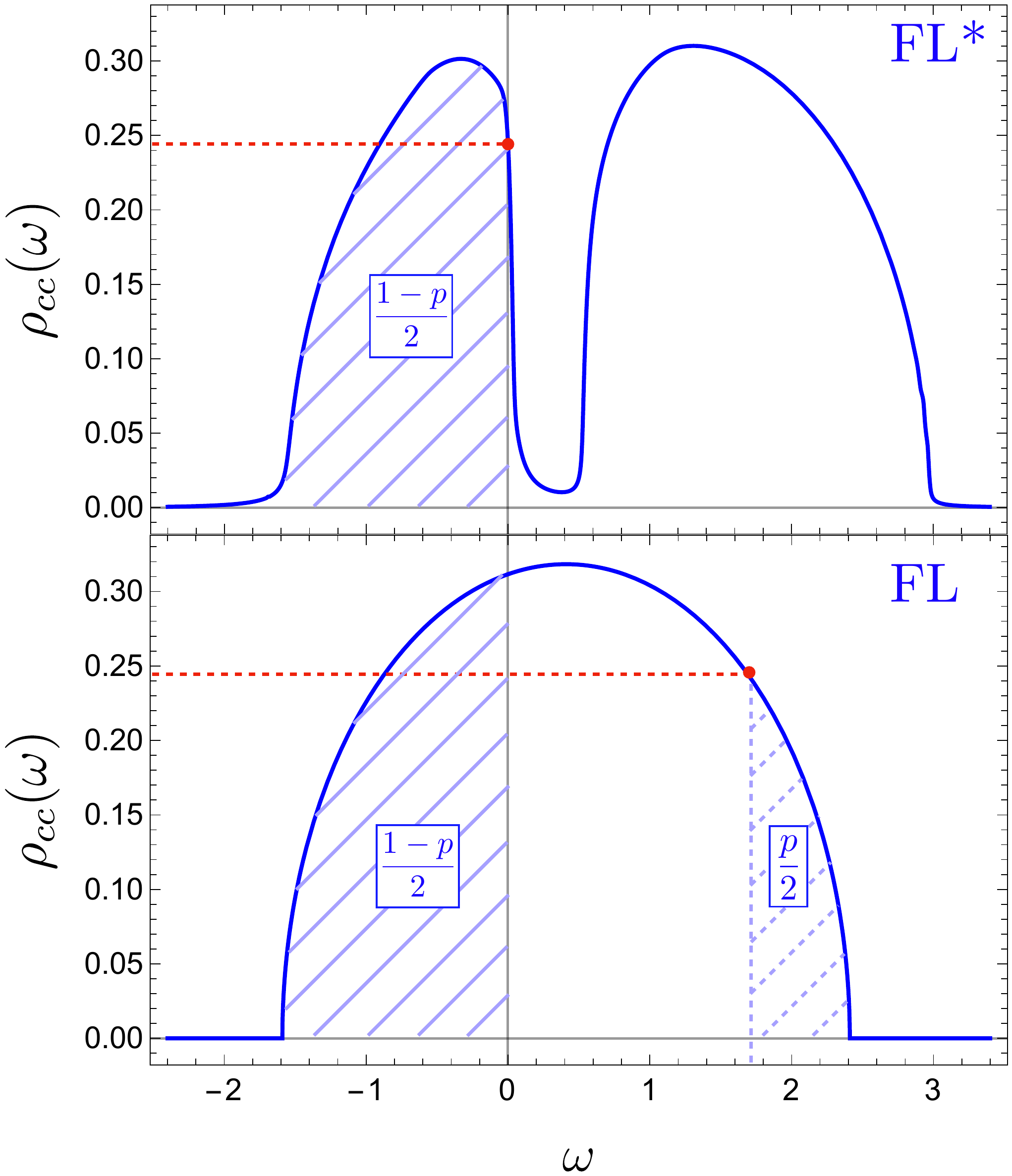}
\caption{Local electron spectral function $\rho_{cc} (\omega) = -(1/\pi) \mbox{Im}\,  G_{cc} (\omega)$ (with $\int_{-\infty}^{\infty} d \omega \rho_{cc} (\omega) = 1$) in the FL* state for the model with random hopping of electrons (these spectra can also be computed, with more numerical effort, for a non-random electron dispersion $\varepsilon_{\bf k}$) showing a particle-hole asymmetric pseudogap near the Fermi level. Also shown is the corresponding Wigner semi-circle spectral function in the FL phase. The areas of the hatched regions are indicated, with $p$ the hole doping away from half-filling. The extended Luttinger theorem in the FL* phase (see Section~\ref{sec:luttinger} and Appendix~\ref{app:luttinger}) implies that the values of $\rho_{cc}(\omega)$ identified by the red circle are equal. The plot is for $p=0.246$, $t=1$, $J=1$, $J_K = 2.03$. }
\label{fig:flsflrhocc}
\end{figure}
In the small Fermi surface FL* phase, we assume that the non-random and antiferromagnetic Kondo coupling $J_K$ dominates, and so the Kondo effect causes the $\Psi_1$ spins to `dissolve' into the Fermi sea of the mobile electrons. By analogy with the corresponding process in the two-band Kondo lattice model, we conclude that the Fermi surface will correspond to an electron density of $1+(1-p) = 2-p$: this is a small Fermi surface of holes of density $p$.  There is an interesting inversion here that is worth noting: at the mean-field level, the {\it small\/} Fermi surface FL* phase of the single-band+ancilla model maps on to the {\it large\/} Fermi surface phase of the two-band Kondo lattice model, where it is the FL phase of that model. At small doping $p$, we can refine this to the statement that the FL* phase of the single band model maps onto a lightly-doped Kondo insulator in a Kondo-Heisenberg lattice model.
This correspondence, however, does not hold beyond mean-field: in the small Fermi surface FL* phase of the single-band+ancilla model there are fractionalized spinon excitations arising from the $\Psi_2$ ancillas, which are required by the generalized Luttinger theorem. There are no fractionalized excitations in the large Fermi surface FL phase of the two-band Kondo lattice model.

(For completeness, we clarify what we mean by `fractionalization' in a metal. Although a FL phase has half-integer spin excitations, it is not fractionalized: all half-integer spin excitations carry an odd electronic charge. A fractionalized excitation has half-integer spin with even charge, or integer spin with odd charge; the zero charge case is the spinon.)

At this point, it is useful to contrast the ancilla qubit approach to the pseudogap phase of the single band cuprates from earlier variational wavefunction approaches. In the popular and influential `vanilla' approach to resonating valence bond theory \cite{Vanilla}, the underdoped normal state is modeled as a Gutzwiller projected Fermi liquid
\bea
\left| \Phi_{\rm vanilla} \right\rangle &=& \bigl[ \mbox{Project out doubly-occupied sites} \bigr] \nn &~&~~~~
\bowtie \bigl| \mbox{Slater determinant of C} \bigr\rangle\,.
\label{vanillawavefunction}
\eea
However, this approach predicts a large Fermi surface for small $p$, which appears to be incompatible with recent experiments \cite{CPLT18,Shen19,Ramshaw20}. In our ancilla approach, the corresponding wavefunction of the FL* phase present at small $p$ is \cite{Yahui20a}
\bea
\left| \Phi_{\rm ancilla} \right\rangle &=&  \bigl[\mbox{Projection onto rung singlets of $\Psi, \widetilde{\Psi}$} \bigr] \nn
&~&~~~~\bowtie \bigl| \mbox{Slater determinant of $(C,\Psi)$} \bigr\rangle \nn
&~&~~~~~~~~~\otimes \bigl| \mbox{Slater determinant of $\widetilde{\Psi}$} \bigr\rangle \label{ancillawavefunction}
\eea
Note that after the projection onto rung singlets on the right hand side of Eq.~(\ref{ancillawavefunction}), $\left| \Phi_{\rm ancilla} \right\rangle$ is a wavefunction dependent only upon the physical $C$ degrees of freedom in the single band model under consideration. 
As has been well understood \cite{LeeWen06} for some time, for $\left| \Phi_{\rm vanilla} \right\rangle$ the consequences of the projection can be understood by examining a gauge theory for the constrained subspace. The gauge symmetry is fully higgsed in the phase described by $\left| \Phi_{\rm vanilla} \right\rangle$, and so gauge fluctuations are not singular; in the language of wavefunctions, the projection in Eq.~(\ref{vanillawavefunction}) has little influence on the large Fermi surface of $C$, apart from a Brinkman-Rice renormalization \cite{BrinkmanRice} of the quasiparticle mass. Turning to $\left| \Phi_{\rm ancilla} \right\rangle$, the implementation of the rung singlet projection requires consideration of a SU(2)$_S$ gauge field \cite{Yahui20a,Yahui20b}: this gauge field is also fully higgsed in the FL* phase, and so the small Fermi surface formed by the Slater determinant of $(C,\Psi)$ in Eq.~(\ref{ancillawavefunction}) is stable under the projection.

With our use of SYK models for the couplings within each ancilla layer, it becomes possible to obtain exact results for the electron spectral function in the FL* phase. A typical spectrum in shown in Fig.~\ref{fig:flsflrhocc}.
This spectrum is for a random matrix model of hopping within the electronic $C$ layer, which leads to a semi-circular density of states in FL phase. We shall show below that computations are also possible for arbitrary translationally invariant band structures within the $C$ layer. As shown in the figure, our numerical results obey a modified Luttinger theorem which constrains the value of the density of states at the Fermi level in the FL* phase. 
A notable feature of Fig.~\ref{fig:flsflrhocc} is the appearance of a pseudogap in the electronic spectrum of the FL* phase. This pseudogap leads to a particle-hole asymmetry in the electron spectral function which has similarities to that discussed in Ref.~\cite{Randeria05}. Moreover, the minimum in the local density of states is slightly above the Fermi level, as is observed in STM experiments at higher temperatures \cite{LeeDavis09}.

It is interesting to ask if there is any holographic parallel, along the lines of Ref.~\cite{SS10}, to the small-to-large Fermi surface transition we describe here. The physical electronic layer has a $q=2$ SYK model with quasiparticle excitations which has no black hole dual, but the question becomes better defined if we consider the non-Fermi liquid phases of higher $q$ analogs of our model. In this case there is a correspondence to wormhole transitions of black holes  \cite{Sahoo:2020unu}, as we will discuss further at the end of Section~\ref{sec:conc}. 

We will define the model and obtain its saddle-point equations in Section~\ref{sec:model}, with details of the $G$-$\Sigma$ theory appearing in Appendix~\ref{app:saddle}. We will discuss some exact features of the phases of the model, including the Luttinger theorems which apply in the FL* and FL phases in Section~\ref{sec:luttinger} and Appendix~\ref{app:luttinger}. Our numerical solutions of the imaginary frequency saddle-point equations and the mean field phase diagram are described in Section~\ref{sec:numerical}. In Section~\ref{sec:spectral} we turn to a discussion of the electronic spectrum in the FL* phase by a direct solution of the saddle point equations for real frequencies. This reveals interesting structure in the spectrum which would have been difficult to obtain by numerical analytic continuation of the imaginary frequency solution. This real frequency analysis is aided by an exact solution of the saddle point equations at $J=0$ which is presented in Appendix~\ref{app:J0}.

\section{Model and saddle-point equations}
\label{sec:model}

We extend the model of Burdin {\it et al\/} \cite{Burdin_2002} to include 2 layers of ancilla as illustrated in Fig.~\ref{fig:layers} to obtain the following Hamiltonian
\begin{align}
H=- & \mu \sum_{i} C^\dagger_{i;\alpha} C_{j;\alpha} + \frac{1}{\sqrt{N}} \sum_{i\neq j} t_{ij} C^\dagger_{i;\alpha} C_{j;\alpha}+\frac{J_K}{M} \sum_i  C^\dagger_{i;\alpha} C_{i;\beta} S_{i;1;\beta \alpha} \notag \\
& +\frac{1}{\sqrt{MN}} \sum_{a=1,2} \sum_{i < j} J_{a;ij}S_{i;a;\alpha \beta} S_{j;a;\beta \alpha}+\frac{J_\perp}{M} \sum_i  S_{i;1; \alpha \beta} S_{i;2;\beta \alpha}\,.
\label{eq:initial_Hamiltonian}
\end{align}
The physical layer has electrons $C_{i;\alpha}$ on the sites $i$ of a lattice, and $\alpha,\beta=1,2,...,M$ are SU($M$) spin indices which generalize the SU(2) spin indices.
The ancilla layers $a = 1,2$ are represented by SU($M$) spins $S_{i;a;\alpha \beta}$. 

We will now take the large spatial dimension and large $M$ limit of (\ref{eq:initial_Hamiltonian}) as discussed by Burdin {\it et al\/} \cite{Burdin_2002}. This can be performed in a model with non-random $t_{ij}$ and a corresponding momentum space dispersion $\varepsilon_{\bf k}$ of the bare electrons $C_{{\bf k}; \alpha}$; such a model will have sharp Fermi surfaces in momentum space, and so a well-defined concept of the `size' of the Fermi surface. However, it is technically somewhat easier to work in a model in which the $t_{ij}$ are independent random numbers representing all-to-all hopping on a cluster of sites $i=1 \dots N$; such a random model has the same phases in the large $M$ limit, and is also constrained by a Luttinger theorem, as we will review in Section~\ref{sec:luttinger}. We will present our analysis for the case of random $t_{ij}$, but will indicate in Section~\ref{sec:luttinger} the modifications needed for the case with non-random hopping and a sharp momentum space dispersion $\varepsilon_{\bf k}$. For the random case, 
the couplings in (\ref{eq:initial_Hamiltonian}) obey
\begin{align}
  &\overline{t_{ij}}=0 \ \ \overline{|t_{ij}|^2}=t^2 \notag\\
  &\overline{J_{1;ij}}=0 \ \ \overline{J^2_{1;ij}}= J^2 \notag\\
  &\overline{J_{2;ij}}=0 \ \ \overline{J^2_{2;ij}}= J^2 
\end{align}
The Kondo exchange coupling $J_K$ and the rung exchange coupling between the ancilla, $J_K$ are taken to be positive ({\it i.e.\/} antiferromagnetic) and non-random.
 
The large $M$ limit is implemented by a fermionic parton representation of the spins with
\begin{align}
S_{i;1;\alpha \beta}&=\Psi^\dagger_{i;1;\alpha}\Psi_{i;1;\beta} \notag\\
S_{i;2;\alpha \beta}&= \Psi^\dagger_{i;2;\alpha} \Psi_{i;2;\beta} 
\label{eq:abrikov_fermion}
\end{align}
where the fermions $\Psi_{1,2}$ obey the local constraint
\begin{align}
\sum_{\alpha} \Psi^\dagger_{i;a;\alpha} \Psi_{i;a;\alpha}&= \frac{M}{2}\,. \label{constraint1}
\end{align}
The chemical potential $\mu$ is adjusted so that the average density of the electrons is
\begin{equation}
\sum_{\alpha}  \left\langle C^\dagger_{i;\alpha} C_{i;\alpha} \right\rangle= \frac{M}{2} (1 - p) \label{constraint2}
\end{equation}

We now discuss the crucial issue of the gauge symmetries introduced by the parton representation in Eq.~(\ref{eq:abrikov_fermion}). 
As written, the model (\ref{eq:initial_Hamiltonian}) has a global U(1) symmetry of the conservation of the electron number $C^\dagger C$, and a pair of U(1) gauge symmetries, denoted U(1)$_1$ and U(1)$_2$, associated with the constraints (\ref{constraint2}) on the two ancilla layers. However, for the theory with the ancillas to apply to the underlying single band Hubbard model, we also need to project the ancilla spins onto the rung singlet subspace. This projection has been discussed at length in earlier papers \cite{Yahui20a,Yahui20b} for $M=2$ (see especially, Section II in Ref.~\cite{Yahui20a}, and Sections II and III.A in Ref.~\cite{Yahui20b}): it was shown that the projection is accomplished by integrating over a rotating reference frame in spin space \cite{Sachdev:2009cp} by employing an additional SU(2)$_S$ gauge symmetry; the subscript $S$ denotes that the {\it spin\/} space rotation, in contrast to the U(1)$_{1,2}$ gauge symmetries which act on the Nambu {\it pseudospin\/} space \cite{MSSS18}. It is possible to extend this SU(2)$_S$ gauge symmetry to the model with general $M$ (similar to Ref.~\cite{RanWen}), but we will not enter into this technical complexity here because it does not change the structure of the large $M$ saddle point. As we have noted earlier, the SU(2)$_S$ gauge symmetry is fully higgsed in the FL* phase, and so does not lead to any singular corrections to the small Fermi surface. In the FL phase, the SU(2)$_S$ gauge fluctuations are confining, and have little influence on the low energy theory of the large Fermi surface.
The SU(2)$_S$ gauge symmetry is important mainly at possible deconfined critical points \cite{Yahui20a,Yahui20b} which we do not address in the present paper, because the transitions are found to be first order at large $M$. 

\subsection{Schwinger-Dyson equations}
\label{sec:sd}

In Appendix~\ref{app:saddle} we describe the formal procedure of taking the large $N$ and large $M$ limits of $H$. 
This procedure yields the following equations for diagonal components of the $3 \times 3$ matrices of Green's functions $G$ and self energies $\Sigma$
\begin{align}
    &\Sigma_{cc}(\tau) = t^2 G_{cc}(\tau) \label{sd1} \\
    &\Sigma_{\psi_1\psi_1}(\tau) = -J^2 G_{\psi_1\psi_1}^2(\tau) G_{\psi_1\psi_1}(-\tau) \label{sd2} \\ 
    &\Sigma_{\psi_2\psi_2}(\tau)= -J^2 G_{\psi_2\psi_2}^2(\tau) G_{\psi_2\psi_2}(-\tau) \label{sd3}
\end{align}
These are just the self-energies of the $q=2$ and $q=4$ complex SYK models. 

The off-diagonal self-energies play an important role in our analysis.
As the interband couplings are non-random, these self-energies are independent of frequency, and we use a different symbol, $R$ for their constant values (similar to Ref.~\cite{Burdin_2002}). So we write
\begin{align}
    &R_{c\psi_1}  = -J_K G_{c\psi_1}(\tau=0^-) \label{sd4} \\
    &R_{\psi_1\psi_2} = -J_\perp G_{\psi_1\psi_2}(\tau=0^-) \label{sd5} \,.
    \end{align}
Then, the Green's functions are related to self energies by the following matrix Dyson equation
in Matsubara frequency $\omega_n$
\begin{align}
    G_{v} (i \omega_n) = - \begin{pmatrix} 
    -i\omega_n - \mu + \Sigma_{cc} (i \omega_n) & R_{c\psi_1} & 0 \\
    R_{c \psi_1} & -i\omega_n - \mu_{\psi_1} + \Sigma_{\psi_1\psi_1} (i \omega_n) & R_{\psi_1\psi_2}\\
    0 & R_{\psi_1\psi_2} &-i\omega_n - \mu_{\psi_2} +\Sigma_{\psi_2\psi_2}(i \omega_n)
    \end{pmatrix}^{-1} \label{sd6}
\end{align}
where the subscript of $G_v$ is any of $(cc), (\psi_1\psi_1), (\psi_2\psi_2), (\psi_1\psi_2), (c\psi_1)$. We have made a gauge choice in which the $R$ are real.

Our task is now to solve equations (\ref{sd1}-\ref{sd6}) for the Green's functions and self energies, where the chemical potentials $\mu$, $\mu_{\psi_1}$, $\mu_{\psi_2}$
are chosen to satisfy (\ref{constraint1},\ref{constraint2}). If there is more than one solution, we have to choose the one with the lowest free energy, expressions for which are presented in Appendix~\ref{app:saddle}.

\section{Phases and the Luttinger relations}
\label{sec:luttinger}

The nature of the phases of the model of Section~\ref{sec:model} are controlled by the values of the real saddle-point variables $R_{c\psi_1}$ and $R_{\psi_1,\psi_2}$. This becomes clear upon examining their role as Higgs fields under the gauge symmetries which were discussed at the end of Section~\ref{sec:model}.

The fields associated with the mean values $R_{c \psi_1}$ and $R_{\psi_1,\psi_2}$ carry charges of the U(1) gauge fields as follows
\beq
\begin{tabular}{c|c|c}
 & ~$R_{c\psi_1}$~ & ~$R_{\psi_1 \psi_2}$~\\
\hline
U(1) & +1 & 0 \\
U(1)$_1$ & -1 & +1 \\
U(1)$_2$ & 0 & -1
\end{tabular} \label{gtable}
\eeq
In the full theory, beyond the large $M$ saddle-point, with projection onto the rung singlet subspace, 
we have to consider a SU(2)$_S$ gauge symmetry, and the fields analogous to $R_{c \psi_1}$ and $R_{\psi_1 \psi_2}$ also carry SU(2)$_S$ gauge charges \cite{Yahui20a,Yahui20b}.
From the charge assignments in (\ref{gtable}), we can deduce the basic properties of the phases found in our numerical analyses:\\
(A) {\it Large Fermi surface}, FL: $R_{c \psi_1} = 0$, $R_{\psi_1\psi_2} \neq 0$.\\
The non-zero value of $R_{\psi_1\psi_2}$ higgses a diagonal combination of U(1)$_1 \times$U(1)$_2$, but leaves the other diagonal combination unbroken. As we will see below, the spectrum of $\Psi_{1,2}$ fermions is fully gapped in this phase, and so there is no obstacle for the unbroken gauge symmetry to confine. So the structure of this phase is as sketched in Fig.~\ref{fig:layers}b: the $\Psi_1$ and $\Psi_2$ fermions confine in a rung-singlet phase, and the $C$ electrons form a Fermi liquid with a Fermi suface of size $1-p$ electrons. The SU(2)$_S$ gauge theory is also confining, but this confinement only influences the already gapped ancilla layers, and has little effect on the large Fermi surface. \\
(B) {\it Small Fermi surface}, FL*: $R_{c \psi_1} \neq 0$, $R_{\psi_1\psi_2} = 0$.\\
Now U(1)$_1$ is higgsed by $R_{c \psi_1}$, and this effectively endows the $\Psi_1$ fermions with the global U(1) charge. The hybridized bands of $C$ and $\Psi_1$ fermions form a Fermi sea of size $2-p$ electrons, which is equivalent to $p$ holes. Indeed, the structure of the state formed by $C$ and $\Psi_1$ is indentical to that obtained by Burdin {\it et al.\/} \cite{Burdin_2002} in their HFL state. The $\Psi_2$ fermions form a gapless $q=4$ SYK spin liquid state with fractionalized spinon excitations, and 
$U(1)_2$ remains unbroken. The SU(2)$_S$ gauge symmetry is fully higgsed by the generalized field analogous to $R_{c\psi}$ \cite{Yahui20a,Yahui20b}, and so the small Fermi surface is stable to SU(2)$_S$ gauge fluctuations. \\
(C) $R_{c \psi_1} \neq 0$, $R_{\psi_1\psi_2} \neq 0$.\\
Both U(1)$_1$ and U(1)$_2$ are now Higgsed, and both the $\Psi_1$ and $\Psi_2$ fermions effectively carry the global U(1) charge; the SU(2)$_S$ gauge symmetry is also higgsed.
This state does appear in our iterative solution of the saddle-point equations. However, upon computation of its free energy, we always find it is metastable, with a free energy higher than the states (A) and (B) above. This state forms a Fermi surface of $3-p$ electrons; subtracting a filled band, this is equivalent to a Fermi surface of $1-p$ electrons. So by Higgs-confinement continuity, we can assume this state is formally the same as the FL phase (A). But both layers of ancilla are involved in the band structure, and so this phase may not be a realistic description of the FL phase of the single-band model. 

Let us now describe the structure of the Luttinger relations in these phases, following earlier work \cite{GeorgesRMP,PG98,Burdin_2000,Burdin_2002,Powell05,Coleman05,GKST,Shackleton_2020}. We note that these relations are expected to be exact, and do not rely upon the large $M$ limit. In the present formulation, we will see that the generalized Luttinger relations obtained earlier by topological arguments \cite{Senthil_2003,SVS04,Paramekanti_2004,Else2020} can also be obtained in a more conventional Luttinger-Ward formalism.

It is useful to first solve the equations in Section~\ref{sec:sd} for $G_{cc}$ and $\Sigma_{cc}$, in terms of the other unknowns. We solve (\ref{sd6}) in the form a continued fraction by writing
\bea
G_{cc} (i \omega_n) &=& \frac{1}{i \omega_n +\mu - \Sigma_{cc} (i \omega_n) - R_{c \psi_1}^2 \mathcal{G}_{\psi_1} (i \omega_n)} \label{sd7} \\
\mathcal{G}_{\psi_1} (i \omega_n) & \equiv & \frac{1}{i \omega_n + \mu_{\psi_1} - \Sigma_{\psi_1 \psi_1} (i \omega_n) - R_{\psi_1\psi_2}^2 \mathcal{G}_{\psi_2} (i \omega_n)} \label{sd8} \\
\mathcal{G}_{\psi_2} (i \omega_n) & \equiv & \frac{1}{i \omega_n + \mu_{\psi_2} - \Sigma_{\psi_2 \psi_2} (i \omega_n)}\,. \label{sd9} 
\eea
Note that $\mathcal{G}_{\psi_1}$ and $\mathcal{G}_{\psi_2}$ are not the same as the Green's functions $G_{\psi_1 \psi_1}$ and $G_{\psi_2, \psi_2}$; rather, they are the Green's functions for $\Psi_1$ and $\Psi_2$ in absence of mixing between $\Psi_1$ ($\Psi_2$) and $C$ ($\Psi_1$). Indeed, from (\ref{sd7}--\ref{sd9}), we can obtain explicit expressions for the remaining Green's functions of (\ref{sd6}) (these expressions are easy to obtain diagrammatically from the Dyson series)  
\bea
G_{\psi_1 \psi_1} (i \omega_n) &=& \mathcal{G}_{\psi_1} (i \omega_n) + R_{c \psi_1}^2 G_{cc} (i \omega_n) \left[\mathcal{G}_{\psi_1} (i \omega_n) \right]^2 \label{sd10a} \\
G_{\psi_2 \psi_2} (i \omega_n) &=& \mathcal{G}_{\psi_2} (i \omega_n) + R_{ \psi_1 \psi_2}^2 G_{\psi_1 \psi_1} (i \omega_n) \left[\mathcal{G}_{\psi_2} (i \omega_n) \right]^2  \label{sd10b} \\
G_{c\psi_1} (i \omega_n) &=& R_{c \psi_1} \, G_{cc} (i \omega_n) \mathcal{G}_{\psi_1} (\omega_n) \label{sd10c} \\
G_{\psi_1 \psi_2} (i \omega_n) &=& R_{\psi_1 \psi_2} \, G_{\psi_1 \psi_1} (i \omega_n) \mathcal{G}_{\psi_2} (\omega_n) \label{sd10d} \,.
\eea

We now observe that (\ref{sd1}) and (\ref{sd7}) form a pair of coupled equations for $G_{cc}$ and $\Sigma_{cc}$; these equations can be solved analytically in terms of the Green's function $G_c^0$ for the $C$ electrons on their own
\bea
G_{cc} (i \omega_n) &=& G_c^0 (i \omega_n + \mu - R_{c \psi_1}^2 \mathcal{G}_{\psi_1} (i \omega_n)) \label{sd10} \\
G_c^0 (z) 
&=& \int_{-\infty}^{\infty} d\Omega\,\frac{D(\Omega)}{z-\Omega}\,,
\label{sd11}
\eea
where the density of mobile electron states $D(\Omega)$ is given by the Wigner semi-circular distribution for the random $t_{ij}$ model:
\beq
D(\Omega)\,=\,\frac{1}{2\pi t^2}\,\sqrt{4t^2-\Omega^2}\,\,\,,\,\,\,\Omega \in[-2t,+2t] \,, \label{sd12}
\eeq
and $D(\Omega) = 0$ for $|\Omega|>2t$.
As discussed in Refs.~\cite{GeorgesRMP,Burdin_2002}, we can now also present the form of the saddle-point equations if we had chosen a disorder-free $t_{ij}$ with a sharp momentum space dispersion $\varepsilon_{\bf k}$: we simply have to replace (\ref{sd12}) by
\beq
D(\Omega) = \sum_{\bf k} \delta( \Omega - \varepsilon_{\bf k})
\eeq
From these $D(\Omega)$ and (\ref{sd11}), the explicit expressions for $G_c^0 (z)$ for all complex $z$ are:
\beq
G_c^0 (z) = \left\{ \begin{array}{ccc} \displaystyle
 \frac{1}{2t^2}\,\left[z \mp \,\sqrt{z^2-4t^2} \right] & , & \mbox{random $t_{ij}$} \\
\displaystyle \sum_{\bf k} \frac{1}{z- \varepsilon_{\bf k}} & , & \mbox{non-random $t_{ij}$}
\end{array} \right.\,, \label{gc0}
\eeq
where the sign in front of the square root is chosen so that $G_c^0(|z| \rightarrow \infty) = 1/z$.
The conduction electron Green's function is then determined by (\ref{sd10}), while the equations for the $\Psi_1$ and $\Psi_2$ Green's functions and self-energies are given by (\ref{sd2},\ref{sd3},\ref{sd10a},\ref{sd10b}). Also, for the non-random $t_{ij}$ we have the full momentum-dependent Green's function of the physical electrons on the lattice in the FL* phase
\beq
G_{cc} ({\bf k}, i\omega_n) = \frac{1}{i \omega_n + \mu - \varepsilon_{\bf k} - R_{c \psi_1}^2 \mathcal{G}_{\psi_1} (i \omega_n) }\,.
\eeq
So the main approximation in this method is that the influence on the ancilla arises only via a ${\bf k}$-independent (but frequency-dependent) self energy, while a more realistic description of the Fermi surface structure of the pseudogap would have a ${\bf k}$-dependent self energy. Also note that
\beq
G_{cc} (i \omega_n) = \sum_{\bf k} G_{cc} ({\bf k}, i\omega_n)\,.
\eeq

We can now state the Luttinger constraint on the solution of the saddle-point equations. The Luttinger analysis \cite{PG98,Burdin_2000,Burdin_2002,Shackleton_2020} is reviewed and extended in Appendix~\ref{app:luttinger}: it fixes the value of chemical potential at $T=0$ to equal
\beq
\mu = E_F + R_{c \psi_1}^2\, \mathrm{Re}\,\mathcal{G}_{\psi_1} (0)
\label{muval}
\eeq
where the Fermi energy $E_F$ is determined from the free electron density of states as the solution of
\beq
2\int_{-2t}^{E_F} d \Omega \, D(\Omega) \, = \, 
\left\{ 
\begin{array}{ccc}
1-p & , & \mbox{Large Fermi surface, FL} \\
2-p & , & \mbox{Small Fermi surface, FL*}
\end{array}\right.\,. \label{EFval}
\eeq
The remarkable fact is that it is the density of states of the non-interacting electrons which exactly determines the value of $E_F$ for the interacting electron problem. The relationship (\ref{EFval}) is illustrated in Fig.~\ref{fig:flsflrhocc}, where the value of $E_F$ for the FL* phase is determined by the position of the red circle in the lower panel showing the density of states in the FL phase. 

We can also fix the nature of the low frequency behavior of the Green's function in the metallic states. In the FL* phase, we expect that the $C$ and $\Psi_1$ Green's functions will have a finite imaginary part at zero frequency, and so 
\beq 
G_{cc} (\tau) \sim G_{\psi_1 \psi_1} (\tau) \sim 1/\tau
\eeq
at large $|\tau|$ at $T=0$. Using (\ref{sd2}), we can deduce that 
\beq
\mathrm{Im}\, \Sigma_{\psi_1 \psi_1} (\Omega) \sim \Omega^2
\eeq
at small $\Omega$. Then from (\ref{sd7},\ref{sd8}) we obtain $\mathrm{Im}\, \mathcal{G}_{\psi_1} (\Omega) \sim \Omega^2$. Along with (\ref{muval}), we can now obtain the exact density of states of the $C$ electrons at the Fermi level from (\ref{sd10})
\beq\label{DoSEF}
-\frac{1}{\pi} \mathrm{Im}\, G_{cc} (i0^+) = D(E_F)\,.
\eeq
This relationship is also illustrated in Fig.~\ref{fig:flsflrhocc} by the equal values of $\rho_{cc} (\omega)$ at the 2 red circles.

\section{Numerical Results}
\label{sec:numerical}

We turn to a numerical solution of the saddle point equations in Section~\ref{sec:sd}. This section will present a numerical analysis with imaginary time Green's functions. Results obtained by solving the equations directly on the real frequency axis will be presented in Section~\ref{sec:spectral}.

\subsection{Zero doping}
\label{secp0}

In this case we set the chemical potentials to zero, $\mu=\mu_{\psi_1} = \mu_{\psi_2} = 0$, by particle-hole symmetry. The numerical solution of the Schwinger-Dyson equations is based on an iteration procedure, starting from a trial Green's function. After convergence, we inspect the values of the off-diagonal self energies $R_{c\psi_1}$ and $R_{\psi_1 \psi_2}$ to determine the nature of each phase. In some cases there are multiple solutions, and we select the solution with the lowest free energy.  
\begin{figure}[H]
\begin{minipage}[h]{0.45\linewidth}
  \center{\includegraphics[width=1\linewidth]{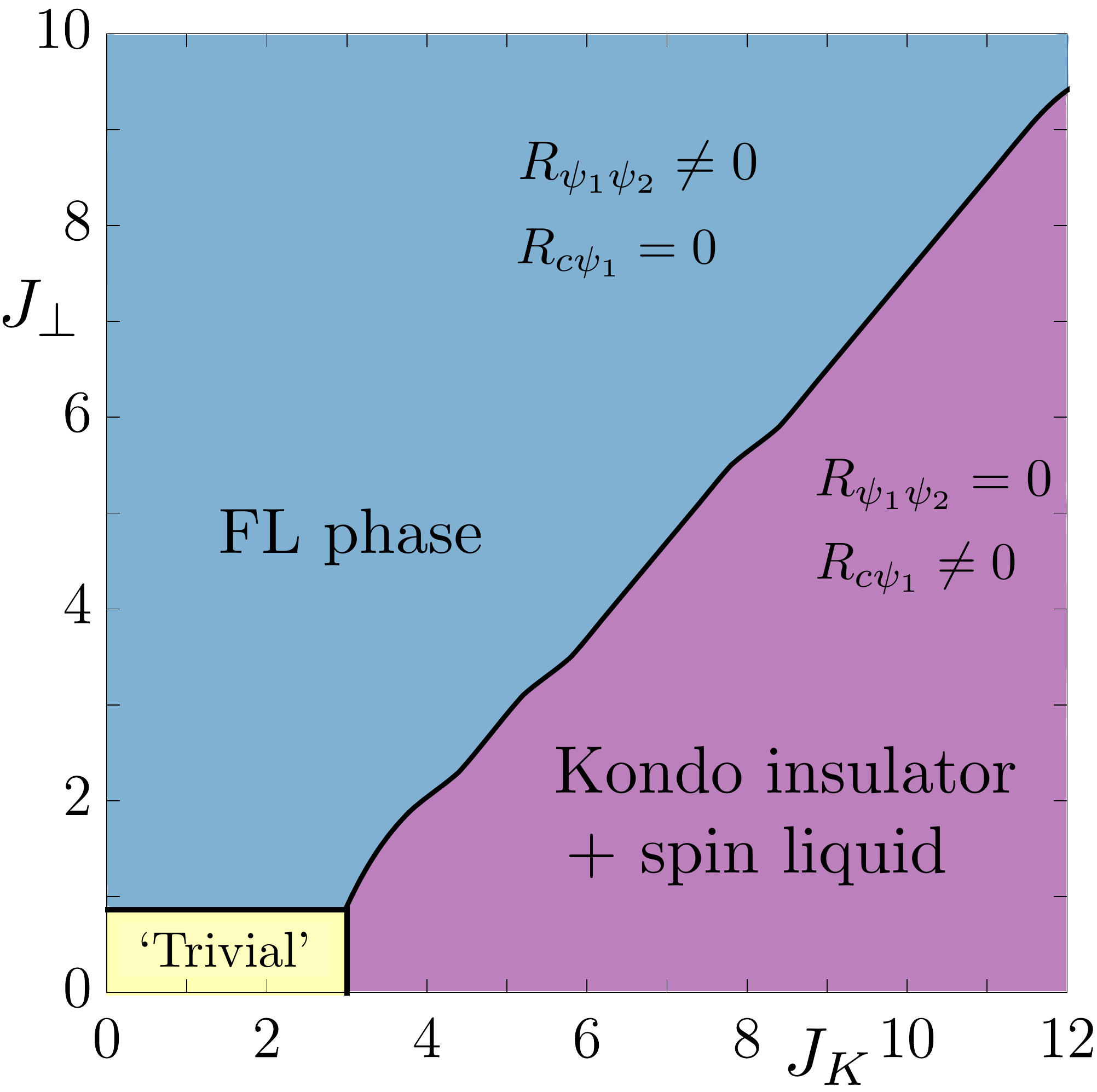}}
  \end{minipage} 
  \hfill    
    \begin{minipage}[h]{0.5\linewidth}  \center{\includegraphics[width=1\linewidth]{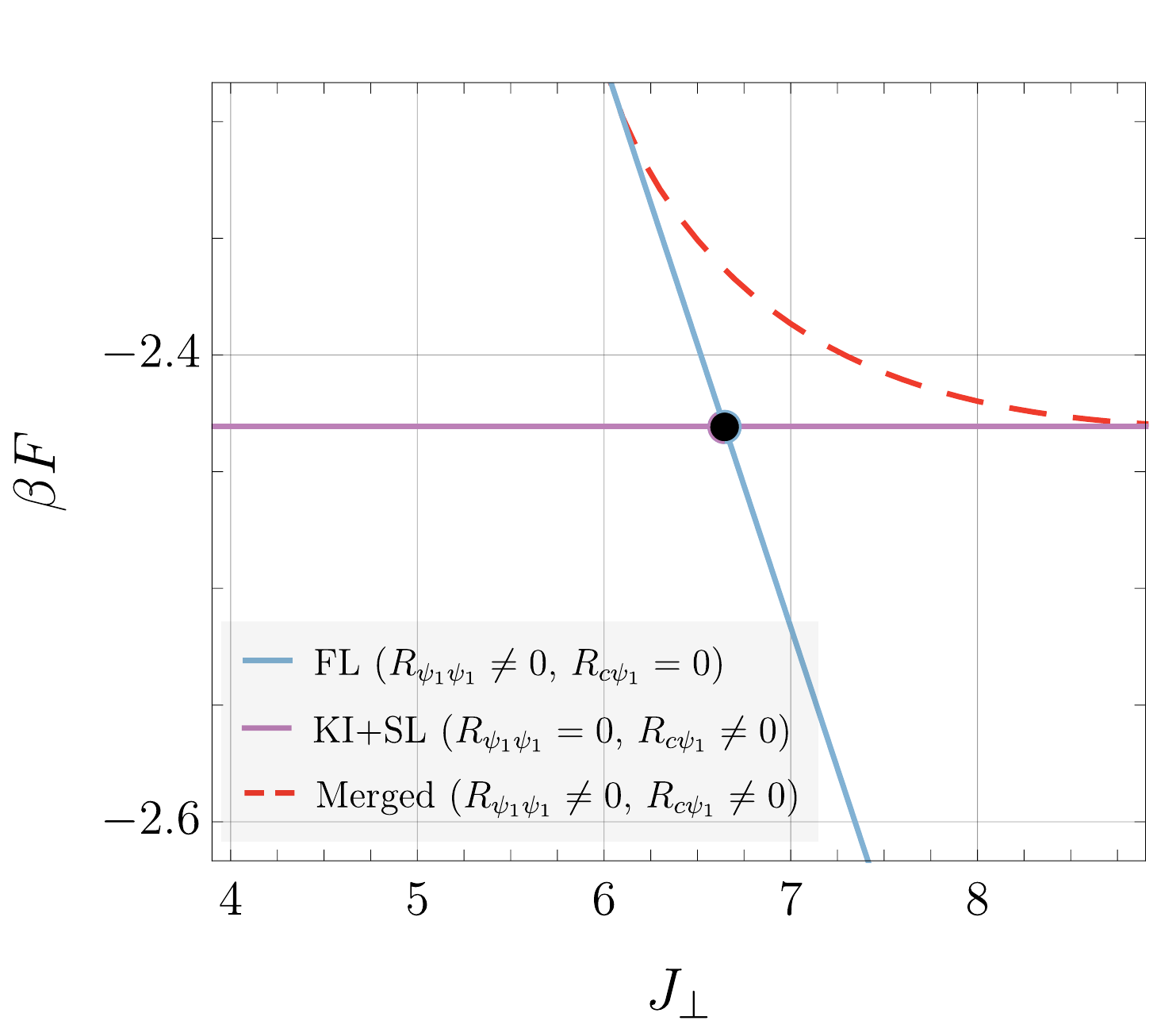}}
    \end{minipage} 
\caption{Phase diagram at zero doping. The right plot shows the free energy of the Fermi liquid (FL), Kondo insulator $+$ spin liquid (KI $+$ SL) and merged phases at $J_K=9$. We note that the merged phase where $R_{c\psi_1} \neq 0$ and $R_{\psi_1\psi_1}\neq 0$ is never dominant. The black dot in the right plot indicates the point where the phase transition happens. Parameters: $t=2$, $J=2$, $\beta \equiv 1/T =100$. }
\label{phase_diag_p0}
\end{figure}

Fig.~\ref{phase_diag_p0}. shows the phase diagram as a function of the two couplings between the layers. When $J_\perp<J_{\perp c}$ and $J_K<J_{K c}$ the `trivial' phase is realized, where all three layers are decoupled,
and $R_{c\psi_1} = R_{\psi_1 \psi_2} = 0$.
The boundaries of this trivial phase can be determined by taking the limit $R_{c \psi_1}\rightarrow 0$ and $R_{ \psi_1\psi_2}\rightarrow 0$ in equations (\ref{sd4},\ref{sd5}): 
\begin{align}
&1=-\frac{J_{Kc}}{\beta}\sum_n G_{cc}^{(d)} G_{\psi_1\psi_1}^{(d)} \label{n1}\\
&1=-\frac{J_{\perp c}}{\beta}\sum_n G_{\psi_2 \psi_2}^{(d)} G_{\psi_1 \psi_1}^{(d)}  \label{n2}\,.
\end{align}
where the superscript $(d)$ implies that the Green's functions are computed with layers decoupled.
Under this condition, the first layer is described by the SYK$_2$ model while second and third layers are described by the SYK$_4$ models (here SYK$_q$ represents the SYK model with random couplings of $q$ fermion operators). It can then been seen for $G_{\psi_1 \psi_1}^{(d)} = G_{\psi_2 \psi_2}^{(d)} \sim 1/\sqrt{\tau}$ that $J_{\perp c}=0$ at $T=0$. So the `trivial' region in Fig.~\ref{phase_diag_p0} shrinks to the line $J_K<J_{K c}$, $J_\perp= 0$ at $T=0$. This `trivial' is expected to be unstable to confinement upon including SU(2)$_S$ gauge fluctuations, and so we will not consider it further. 

In the large $J_K$ regime of Fig.~\ref{phase_diag_p0}, the physical and first ancilla layers are strongly coupled. This implies that the first layer ancilla spins are Kondo screened by the physical electrons. As the latter are at half filling, we realize a Kondo insulator. However, the second ancilla layer remains a SYK$_4$ spin liquid. So while a conventional Kondo insulator is smoothly connected to a trivial band insulator, that is not the case in our model. We instead realize a Mott insulator with fractionalization, with the fractionalized spinon excitations residing on the second ancilla layer. Indeed the presence of the fractionalized excitations in this insulator is required by the extended Luttinger theorem.

In the other limit, when $J_\perp$ is large enough, the free energy of the FL becomes lower. 
Here the physical layer is decoupled from the ancilla layers, and forms a free electron metal. 
On the other hand, the SYK$_4$ spin liquids on the ancilla layers are coupled by $J_\perp$, and this drives them into a gapped state which is smoothly connected to a band insulator. 

As $R_{c \psi_1}$ and $R_{\psi_1 \psi_2}$ change discontinuously between the phases above, the transition between FL and the Kondo insulator is first order. When $J_\perp$ and $J_K$ are close to each other, we also obtain a `merged' solution, with both $R_{c\psi_1}$ and $R_{\psi_1 \psi_2}$ non-zero. However, Fig.~\ref{phase_diag_p0} shows that the free energy of this merged solution is never the global minimum, and the transition is first order.

\begin{figure}[H]
\begin{minipage}[h]{0.45\linewidth}
  \center{\includegraphics[width=1\linewidth]{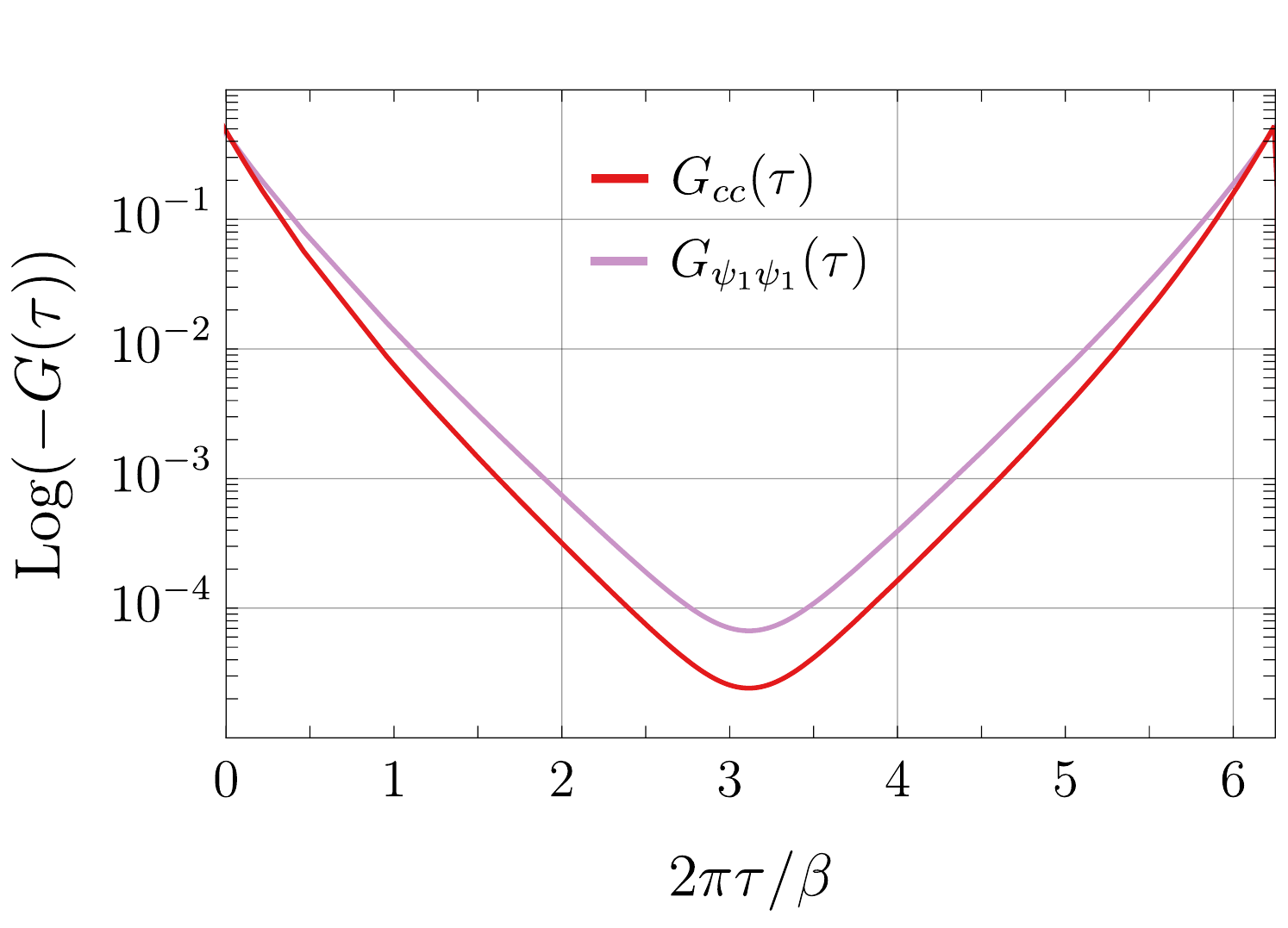}}
  \end{minipage} 
  \hfill    
    \begin{minipage}[h]{0.45\linewidth}
    \center{\includegraphics[width=1\linewidth]{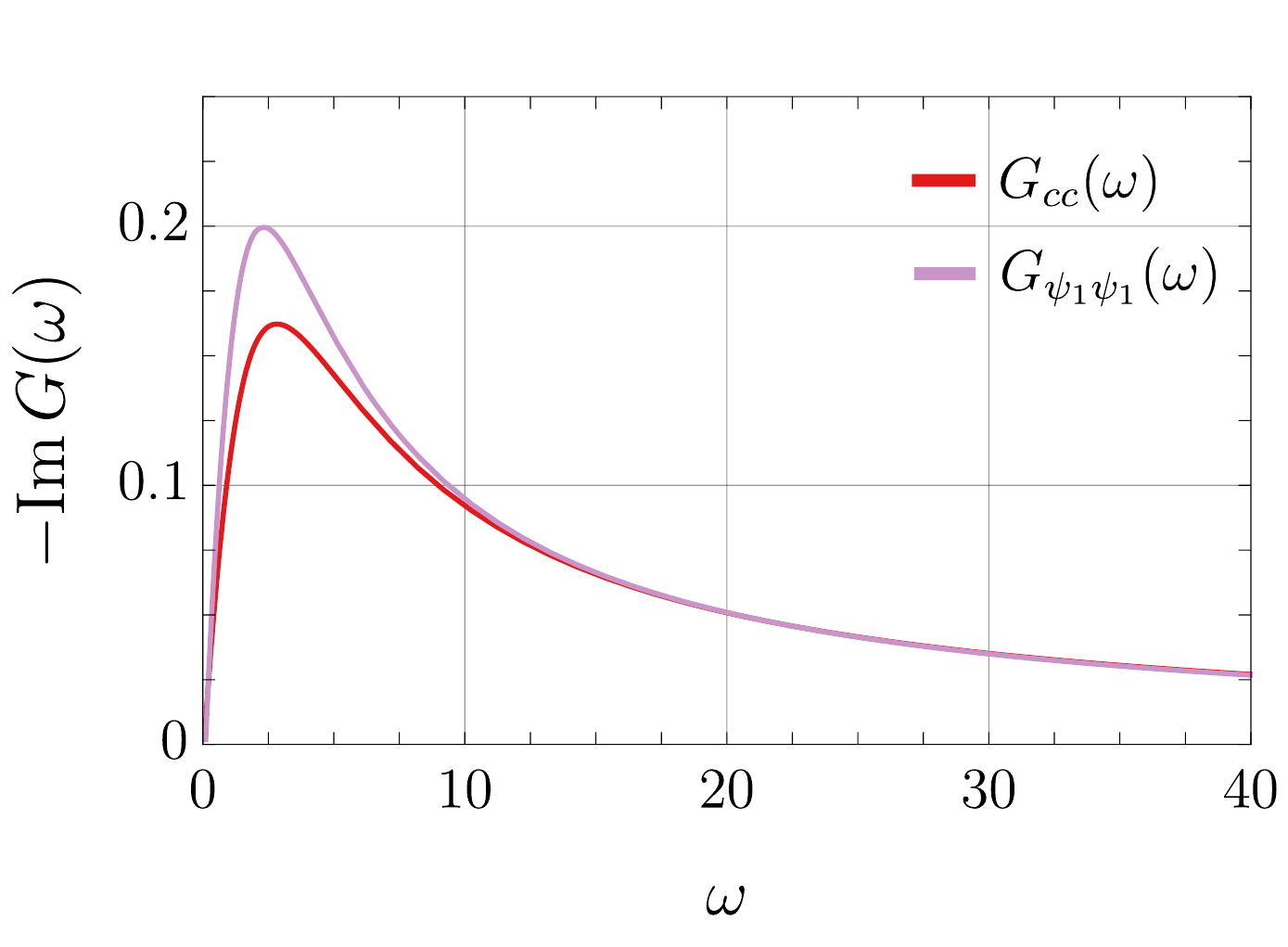}}
    \end{minipage} 
\caption{Imaginary time Green's functions in the Kondo insulator phase at zero doping. The frequency $\omega$ is on the Matsubara frequency axis. Parameters: $t=2$, $J=2$, $J_K=6$, $\beta=10$. }
\label{fig:green_function_p0}
\end{figure}

Fig.~\ref{fig:green_function_p0} shows the Green's functions in the Kondo insulator phase.
We observe exponential decay, indicating the presence of a gap in the physical layer and the first ancilla layer
(there is no gap on the second ancilla layer, which forms a gapless SYK$_4$ spin liquid).
We will obtain a more accurate determination of the gap in the real frequency solution in Section~\ref{sec:spectral}.

\subsection{Non-zero doping}
\label{secpnonzero}

When we turn to non-zero $p$, the Kondo insulator phase in Fig.~\ref{phase_diag_p0} turns into the metallic FL* phase, due to a difference between the density of mobile $C$ electrons and the density of spins in the first ancilla layer. The size of the Fermi surface will be $2-p$, which is equivalent to a density of $p$ mobile holes. 

The density of the electrons in the first layer is equal to $1-p$, while in both ancilla layers it is equal to $1$. This is equivalent to the constraints: $-G_{cc}(\beta^-)=(1-p)/2$ and $-G_{\psi_1 \psi_1}(\beta^-)=-G_{\psi_2 \psi_2}(\beta^-)=1/2$. These can be satisfied by tuning only one chemical potential $\mu$ in the FL phase (while $\mu_{\psi_1}=\mu_{\psi_2} = 0$), and tuning both chemical potentials: $\mu$ and $\mu_{\psi_1}$ in the FL$^*$ phase (while $\mu_{\psi_2}=0$).

\begin{figure}
\includegraphics[width=4in]{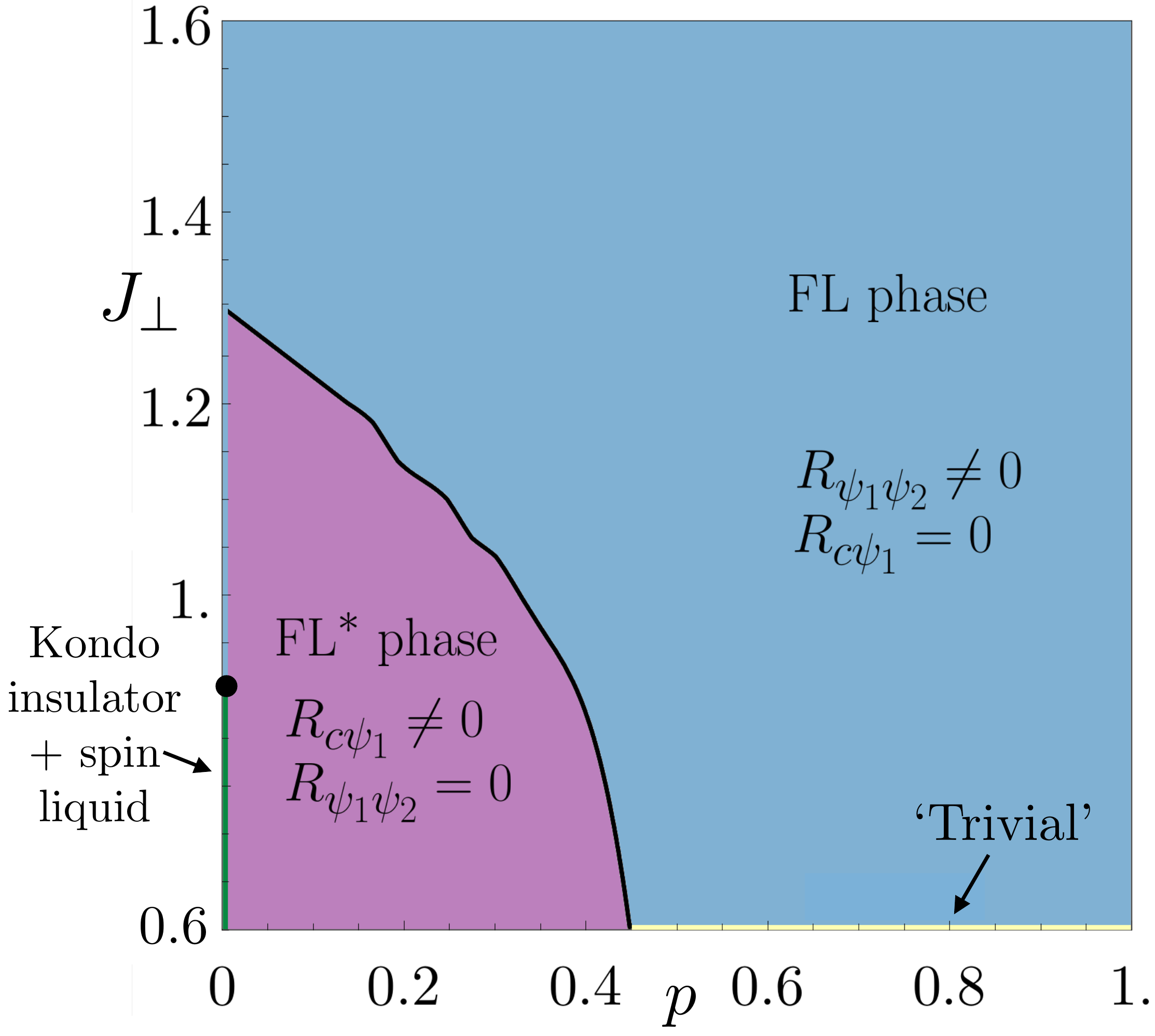}
\caption{Phase diagram as a function $J_\perp$ and $p$. At $p=0$, $J_{\perp c}$ is non-zero because of the non-zero temperature, and the Kondo insulator is present below the black circle. The chemical potential changes discontinuously at any non-zero $p$, and so the phase boundary of the FL* phase does not meet the black circle. Parameters: $t=2$, $J=2$, $J_K = 2.6$,$\beta \equiv 1/T =100$. }
\label{phase_diag_p}
\end{figure}

Derivation of the two-dimensional phase diagram as in the Fig.~\ref{phase_diag_p0} for $p>0$ is complicated because the chemical potentials are unknown. We show a phase diagram as a function of $p$ and $J_\perp$ in Fig.~\ref{phase_diag_p} obtained as described below.

\begin{figure}
 \includegraphics[width=4in]{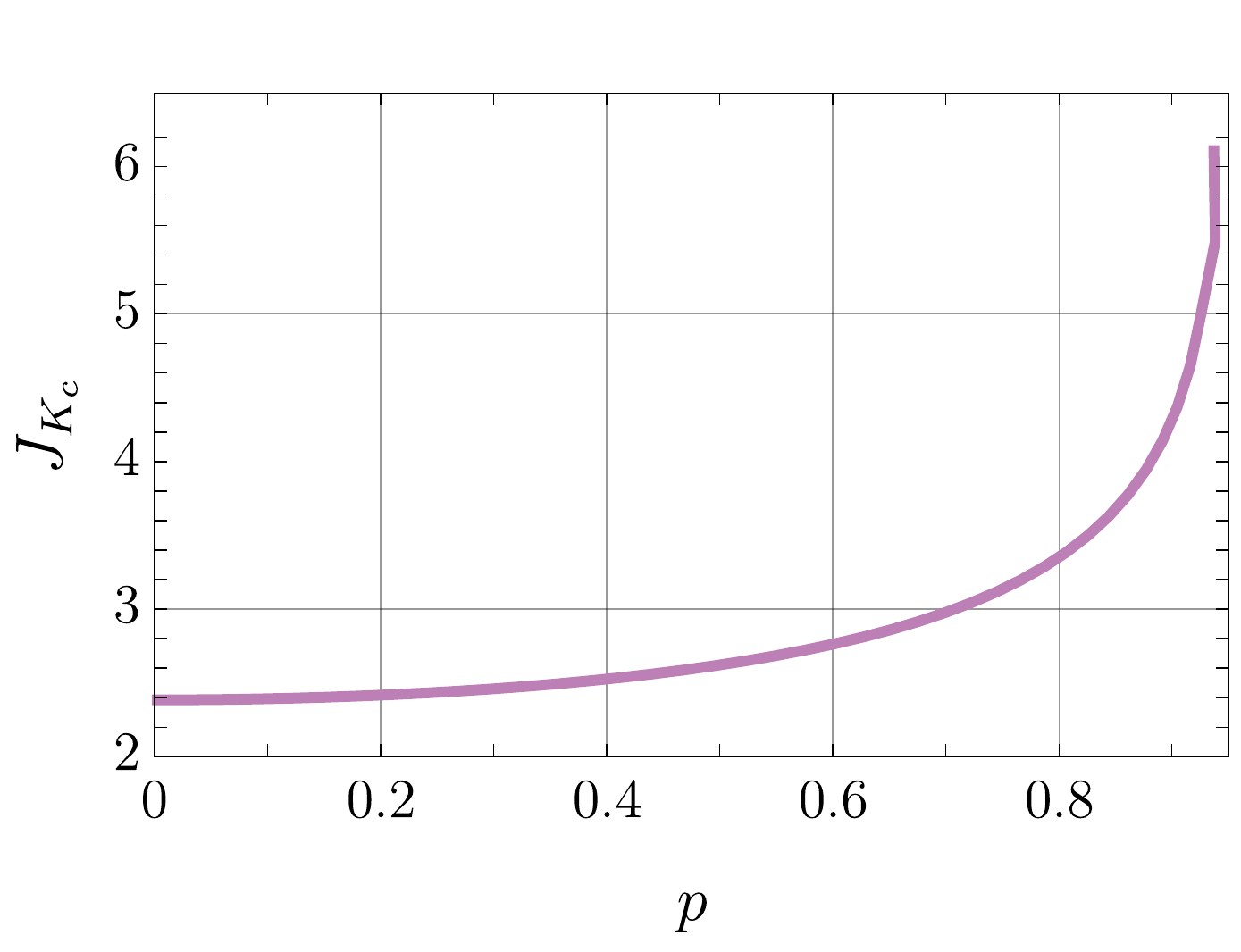}
\caption{The onset of the FL* phase from the trivial decoupled phase. The FL* phase is present for $J_K > J_{K c} (p)$. Parameters: $t=2$, $J=2$, $\beta=100$. }
\label{J_kc_p}
\end{figure}

As in Section~\ref{phase_diag_p0}, the equations (\ref{n1}), (\ref{n2}) can be used to find the values of $J_{Kc}$ and $J_{\perp c}$ that determine the phase transition lines from the trivial phase to FL and FL$^*$ phases. It is clear that $J_{\perp c}$ does not depend on doping and is equal to zero at zero temperature, while $J_{K c}(p)$ depends on doping in a nontrivial way, it increases at larger doping (see Fig.~\ref{J_kc_p}). This has an important physical consequence: the transition between FL and FL$^*$ phases can be initiated by varying doping, without changing any physical couplings. As the doping increases, the onset of the FL$^*$ phase goes to  a larger $J_K$ and the FL phase emerges.

We focus on a fixed $J_K>J_{Kc}$ and show that FL$^*$ phase exists for all dopings, and it goes to the FL phase at large $J_\perp$. As the chemical potentials are unknown, we solve the Schwinger-Dyson equations for all chemical potentials, and impose the constraints afterwards. Our numerical solutions are also aided by analytical solutions which are possible at $J=0$, as described in Appendix~\ref{app:J0}.

 \begin{figure}
\includegraphics[width=0.5\linewidth]{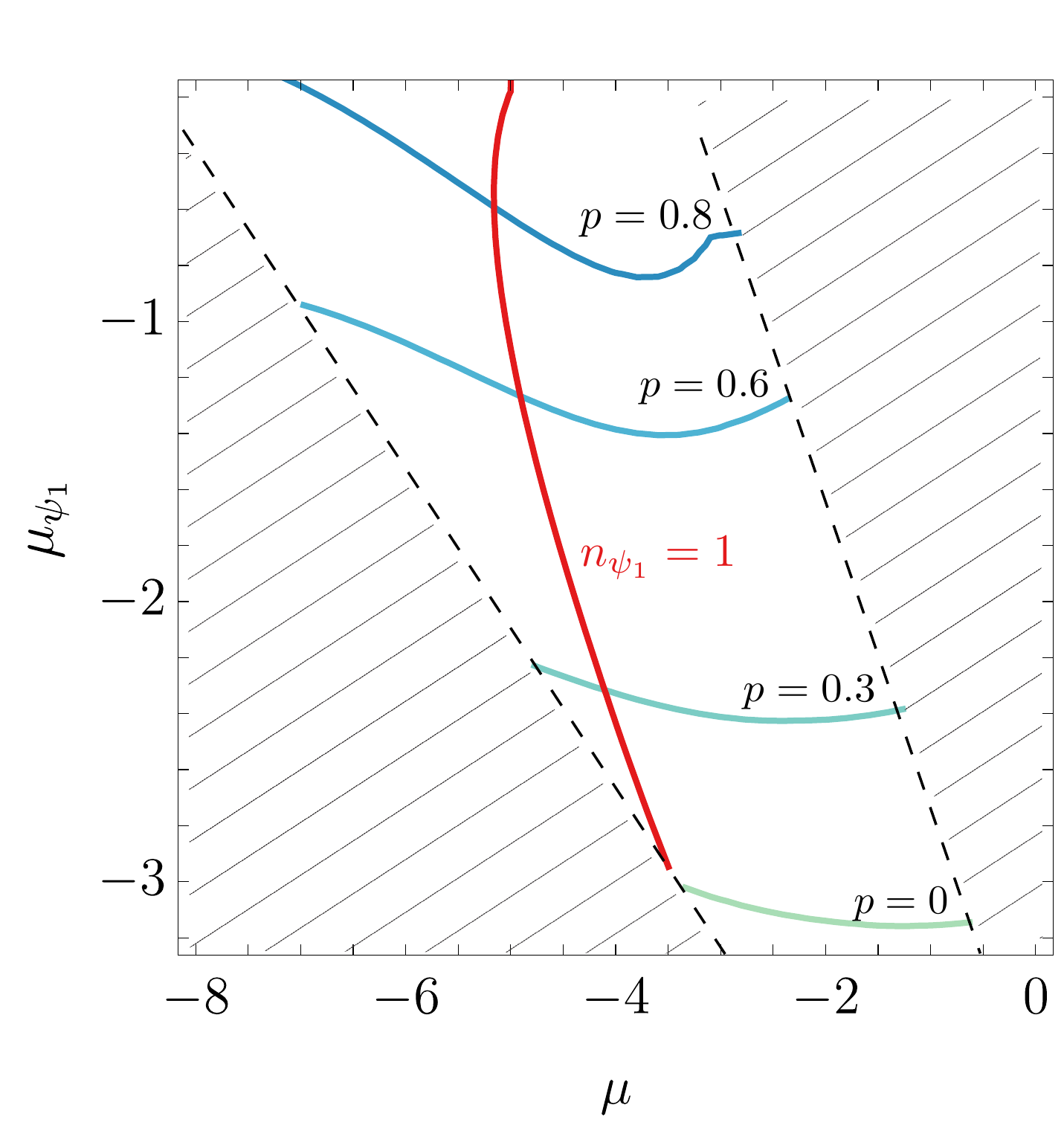}
\caption{Red lines show the contours of constant doping, while blue line shows the contour of constant density in the second layer. The  dashed lines delineate the region of convergence of Schwinger-Dyson equations. The intersection of red line with the lines of constant $p$ gives chemical potentials at the concrete doping. Parameters: $J_K=10$,  $t=2$, $J=2$, $\beta=10$. }
\label{fig:chem_pot}
\end{figure}

Fig.~\ref{fig:chem_pot} displays the lines of the constant densities as functions of two chemical potentials in the FL$^*$ phase. The red line corresponds to a fixed density $n_{\psi_1}=1$ in the first ancilla layer. It intersects with the lines at constant $p$ which indicates the presence of FL$^*$ phase. We note that the line at constant density does not intersect with the $p=0$ line which is expected since the chemical potentials are zero at zero doping. The jump in the chemical potentials from $p=0$ to $p\neq 0$ is consistent with the Luttinger theorem (\ref{muval}). 

\begin{figure}
\includegraphics[width=3.2in]{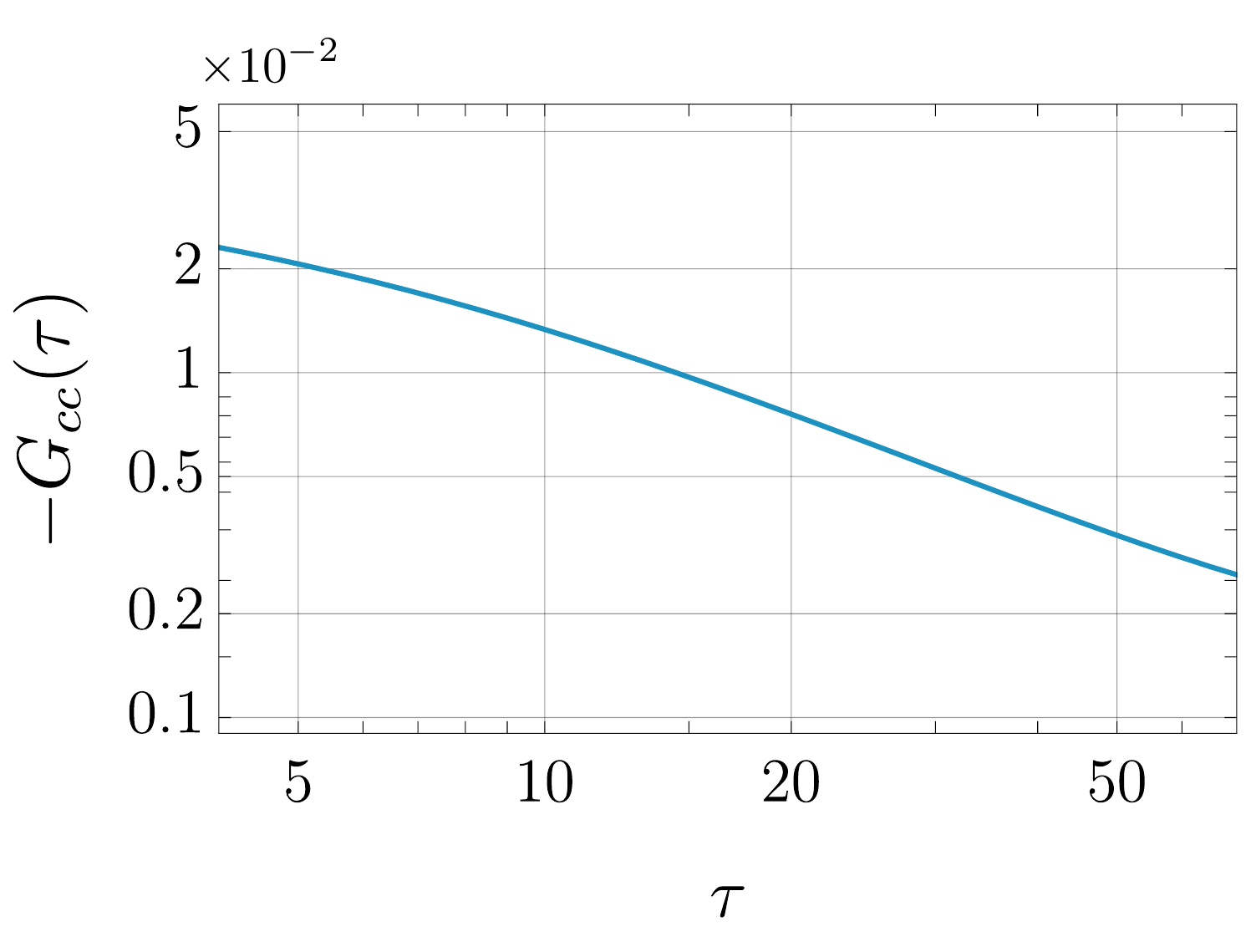}\qquad
\includegraphics[width=3.4in]{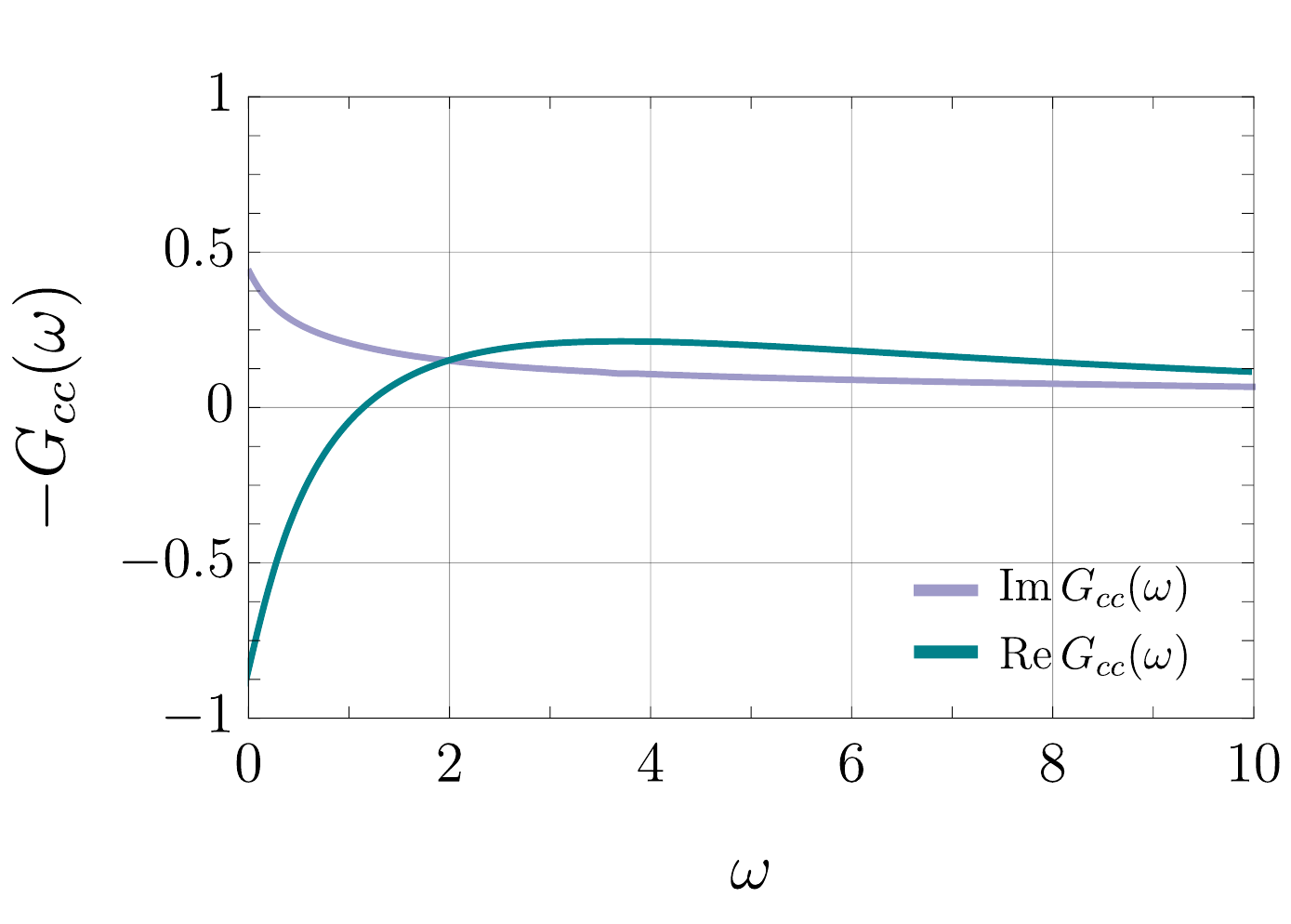}
\caption{Green's functions in FL$^*$ phase at nonzero doping as functions of imaginary time (left, logarithmic scale) and imaginary frequency (right). Parameters: $p=0.4$, $t=2$, $J=2$, $J_K=10$, $\beta=200$. }
\label{fig:green_function_pn0}
\end{figure}

This conclusion can be further substantiated by analyzing the behaviour of Green's functions. Imaginary time Green's functions decay as $1/\tau$ at large times (Fig.~\ref{fig:green_function_pn0}), while in the frequency space they reach constant values at $\omega =0 $ and decay as $1/\omega$ at large frequencies. This demonstrates the gapless metallic nature of the FL* phase. 

\begin{figure}[H]
\begin{minipage}[h]{0.47\linewidth}
  \center{\includegraphics[width=1\linewidth]{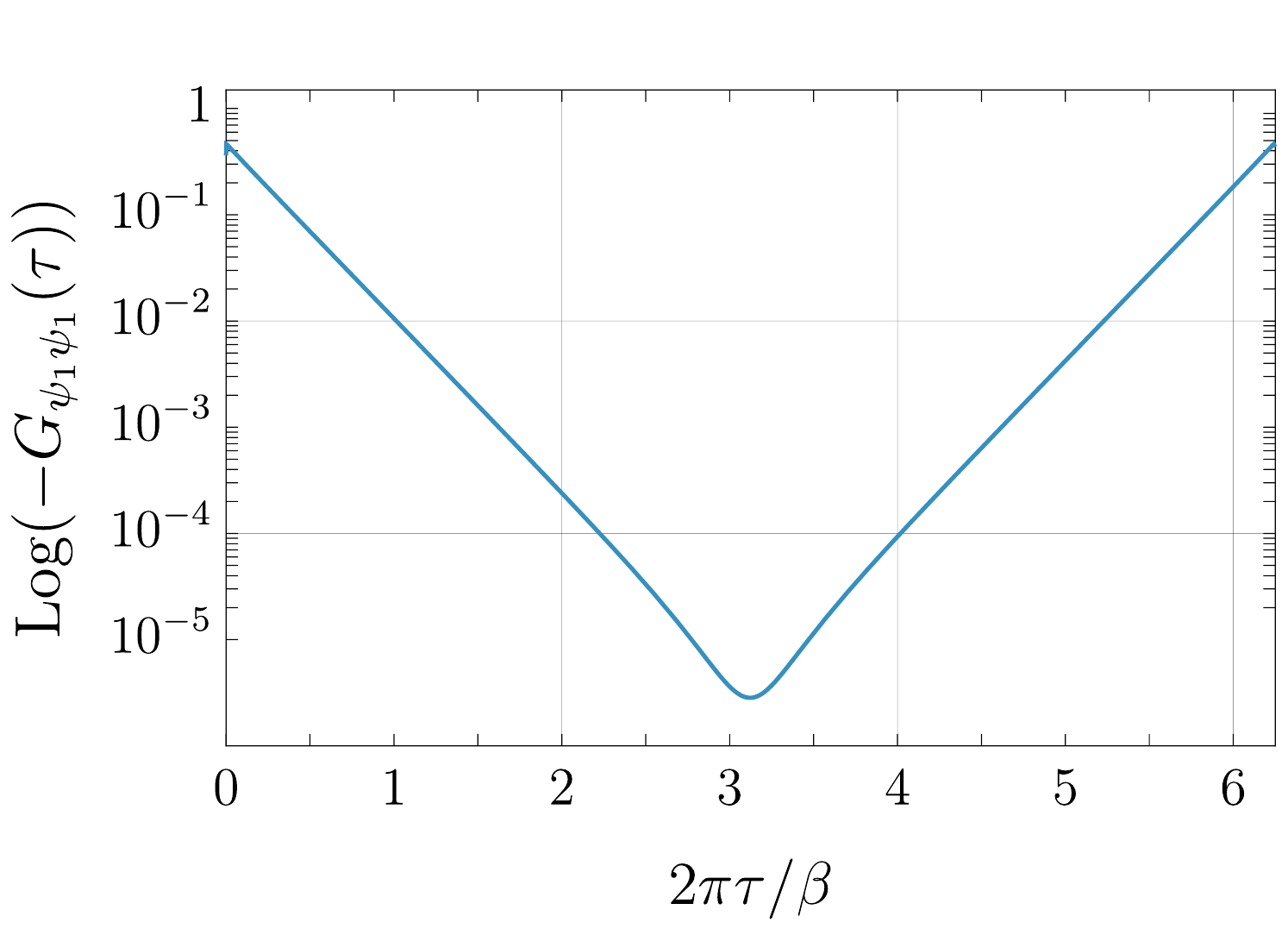}}
  \end{minipage} 
  \hfill    
    \begin{minipage}[h]{0.48\linewidth}
    \center{\includegraphics[width=1\linewidth]{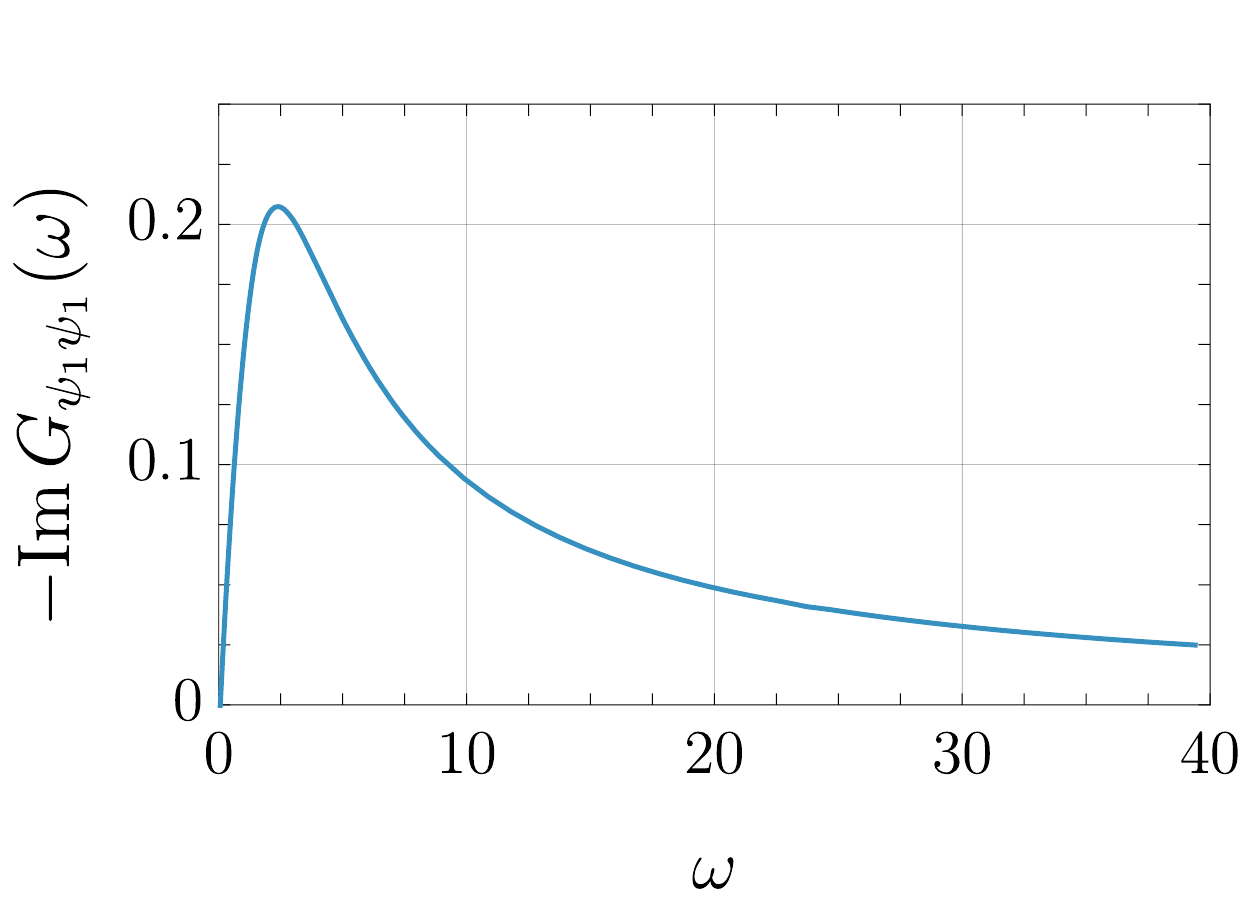}}
    \end{minipage} 
\caption{Imaginary time Green's functions of the two coupled Ancilla layers. The frequency $\omega$ is on the Matsubara frequency axis. Parameters:  $J_\perp=5$, $J=2$, $\beta=10$. }
\label{fig:green_function_fl}
\end{figure}

In the FL phase, the two ancilla layers decouple from the physical layer and form a `trivial' insulator. The decoupled ancilla layers are a pair of the SYK$_4$ models with a non-random exchange coupling $J_\perp$ between them. This is similar, but not identical, to coupled SYK models considered in the literature: there have been studies of SYK models each with a different random 4-fermion term, coupled with another random 4-fermion term \cite{Gu:2016oyy,Chen:2017dbb,Milekhin:2021cou,Milekhin:2021sqd}; and of SYK models each with the same random 4-fermion term, coupled by non-random 2-fermion terms \cite{Maldacena:2018lmt,Gao:2019nyj,Plugge:2020wgc,Sahoo:2020unu,Zhou:2020kxb,Zhou:2020wgh,Zhang:2020szi}. In our case with a non-random 4-fermion coupling $J_{\perp}$, the argument below (\ref{n2}) implies that an infinitesimal $J_{\perp}$ induces a gap. In our numerical study,
the imaginary time Green function demonstrates an exponential decay at large times (Fig.~\ref{fig:green_function_fl}). This implies that the ancilla excitations are gapped, and do not contribute to the low energy excitations of the physical layer.

\section{Spectral functions in the FL* phase}
\label{sec:spectral}



This section solves the saddle-point equations along the real frequency axis, allowing us to obtain accurate results for the electron spectral function. We will consider only the FL* phase, where we have new results on electron spectral functions which do not obey the conventional Luttinger theorem. As the second ancilla layer decouples from the first two layers in the FL* phase, the equations we need to solve are:
\bea
&&\Sigma_{\psi_1\psi_1} (\tau)= -J^2G^2_{\psi_1\psi_1}(\tau)G_{\psi_1\psi_1}(-\tau),\nonumber\\
&&G_{cc} (z) = G_c^0 \left( z + \mu - \frac{R_{c \psi_1}^2}{z + \mu_{\psi_1} - \Sigma_{\psi_1\psi_1}(z)} \right), \\
&&G_{\psi_1\psi_1} (z) = \frac{1}{z+\mu_{\psi_1}-\Sigma_{\psi_1\psi_1}(z)} + \frac{R_{c \psi_1}^2 G_{cc} (z)}{(z + \mu_{\psi_1}-\Sigma_{\psi_1\psi_1}(z))^2}\nonumber\label{Jnon0}
\eea
We solve these equations by iteration.
As before, we fix $R_{c\psi_1}$ to some value, and the chemical potential $\mu_{\psi_1}$ is to be found such that the constraint on the half-filling on the first level \eqref{constraint1} is satisfied.

In both cases of zero and non-zero doping we solve the equations on the spectral functions directly that are related to the Green's functions as follows: 
\begin{align}
G_v(z) =  \int_{-\infty}^{+\infty} d\omega\frac{\rho_v(\omega)}{z-\omega},
\end{align}
where the subscript $v$ indicates either $(cc)$ or $(\psi_1\psi_1)$. The self energies are complex valued, thus we consider equations for both real and imaginary parts. The imaginary part can be obtained directly from the Schwinger-Dyson equation \eqref{Jnon0}. Taking the Fourier transform of the first equation, the imaginary part of the self-energy can be written in the following form
\begin{align}\label{eq:sigma1}
\Sigma''_{\psi_1\psi_1}(\omega>0) &= 
- \pi J^2 \omega^2\int_{0}^{\frac{\pi}{2}} du\sin^3 u \,\,  \rho_{\psi_1\psi_1}(-\omega\cos^2 u)\\
&\times\int_{0}^{\frac{\pi}{2}}d\phi\,\sin^2 2\phi\,\, \rho_{\psi_1\psi_1}(\omega\sin^2u\cos^2\phi)\rho_{\psi_1\psi_1}(\omega \sin^2u\sin^2\phi)\nonumber
 \end{align}
The above expression is defined at positive frequencies. For negative frequencies we change the sign $\omega \to -\omega$. The real part is obtained using the Kramers-Kronig relations. 

To obtain the solutions we are interested in, we choose the exact expressions for the spectral functions $\rho_{cc}(\omega)$ and $\rho_{\psi_1\psi_1}(\omega)$ at $J=0$ (see Appendix~\ref{app:J0}) as the initial functions and proceed with iterations until the needed convergence is reached. 

We use slightly different equations to obtain solutions at zero and non-zero doping. At $p=0$, i.e. when the chemical potentials are set to zero, the equations on the spectral functions are obtained as the imaginary parts of the Green's function in \eqref{Jnon0}.

For the case of $p\neq 0$ we instead use the matrix Dyson equation \eqref{sd6} for the FL$^*$ phase, and consider the imaginary parts of both $G_{cc}(\omega)$ and $G_{\psi_1\psi_1}(\omega)$. We find that these equations converge easier to the solutions that are discussed below. In both cases, we use the same equations for the self-energies \eqref{eq:sigma1}.

As in Section~\ref{sec:numerical}, we consider $p=0$ and $p > 0$ cases in turn.

\subsection{Zero doping}
\label{sec:kondo_ins}

At $p=0$, our FL* phase reduces to Mott insulator. However, the Mott insulating behavior is realized in a novel way in the ancilla approach. As we noted in Section~\ref{secp0}, the electron layer combines with the first ancilla layer to form a Kondo insulator. Then the second ancilla layers realizes a gapless SYK$_4$ spin liquid, which is required to exist in a Mott insulator in a single band model at half filling. In our mean-field analysis, the second ancilla layer decouples in the FL* phase, and we will not consider it further here.

\begin{figure}
\includegraphics[width=5in]{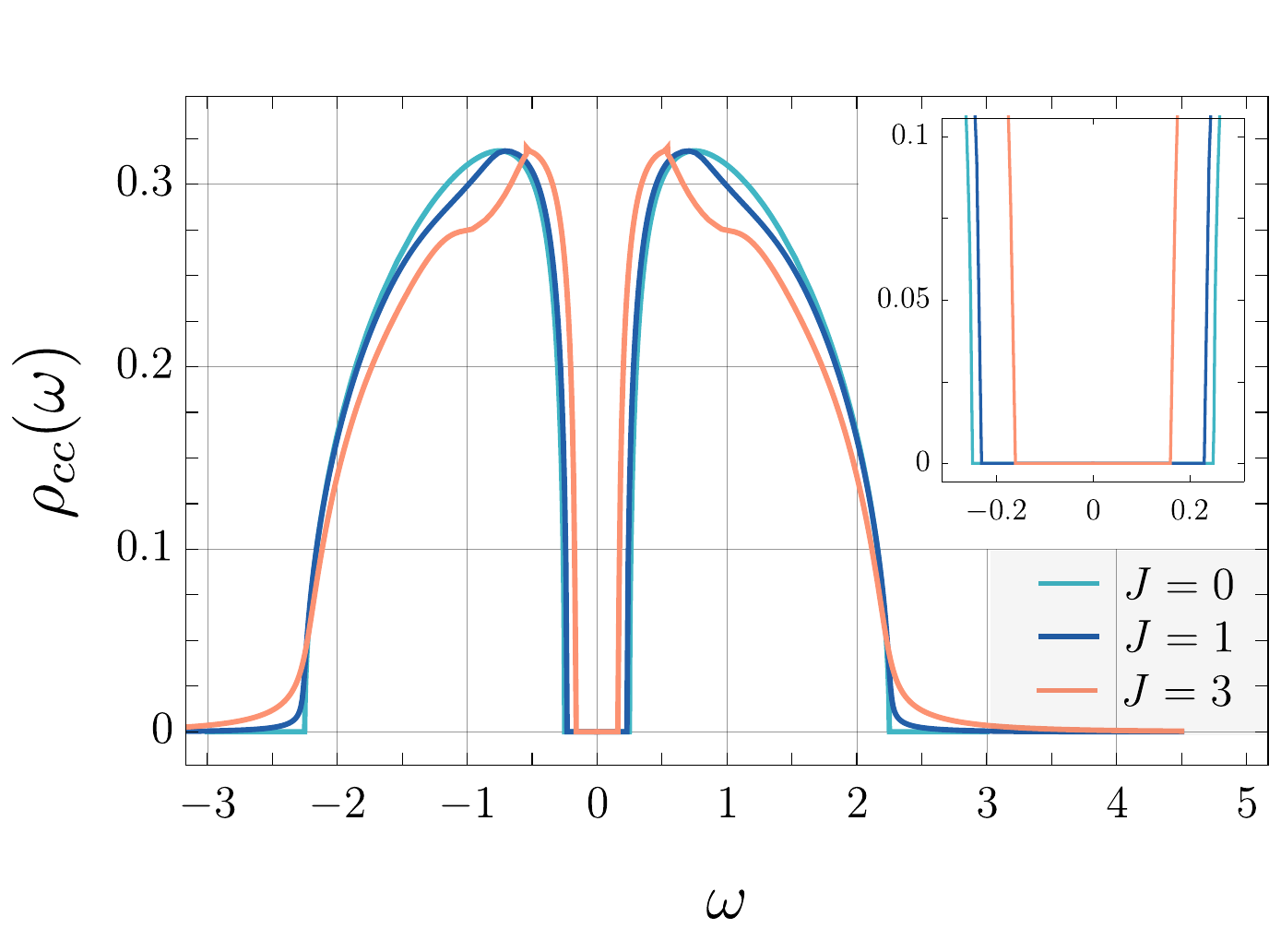}
\caption{Electron spectral density for different values of $J$ and fixed $R_{c\psi_1}=0.75$, in the Mott insulator ({\it i.e.\/} Kondo insulator in the electron and first ancilla layer) at doping $p=0$. Inset: Behavior of the spectral densities at small frequencies. The gap becomes smaller with increasing $J$.}
\label{fig:rhoc_p0_J}
\end{figure}

\begin{figure}
\includegraphics[width=5in]{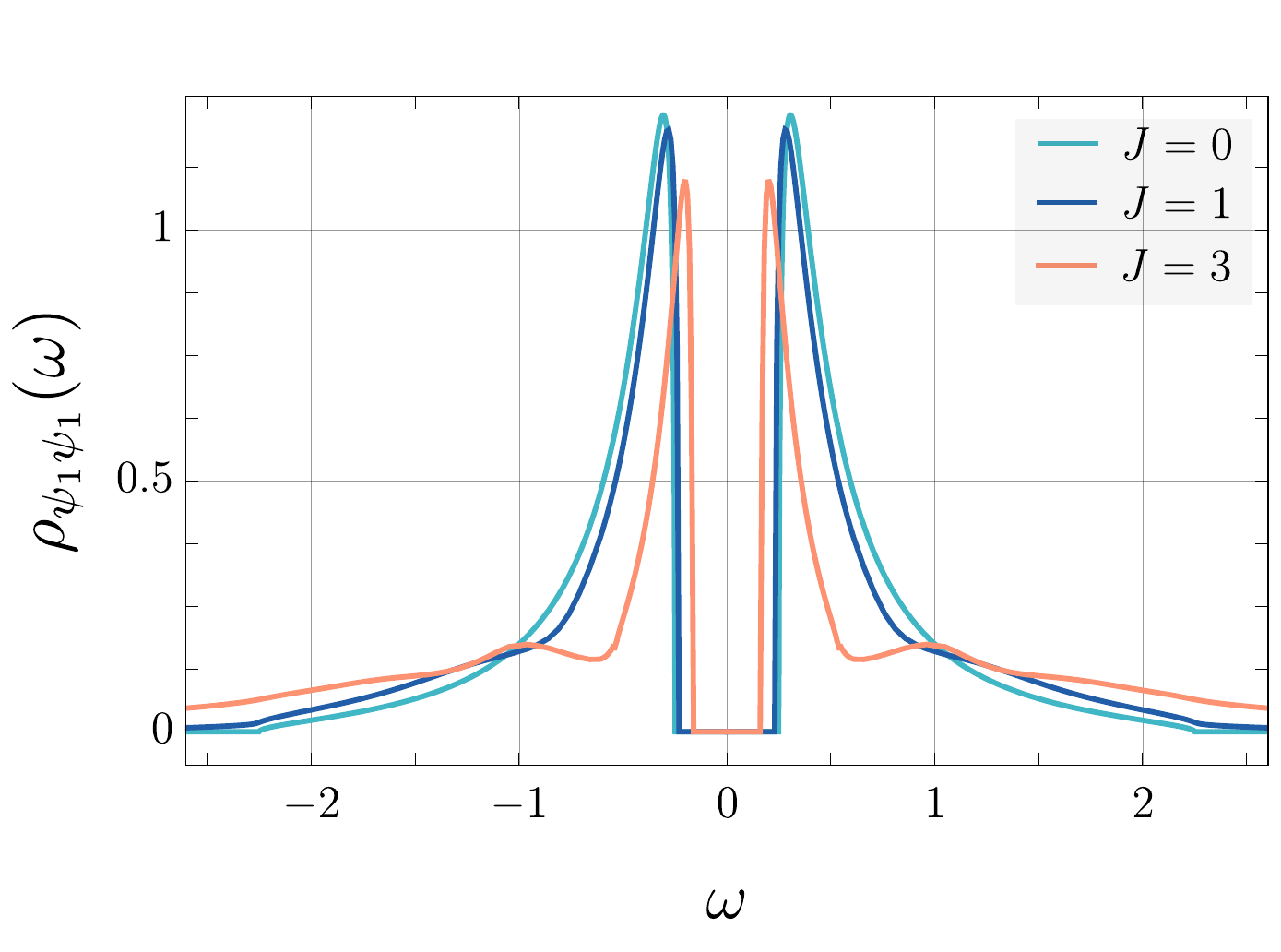}
\caption{Spectral density of $\Psi_1$ fermion for the case in Fig.~\ref{fig:rhoc_p0_J}, with different values of $J$ and fixed $R_{c\psi_1}=0.75$, in the Mott insulator at doping $p=0$.}
\label{fig:rho1_p0_J}
\end{figure}

We define the spectral densities $\rho_v(\omega)=-\text{Im} G_v^R(\omega)/\pi$ and find their behaviors at zero doping $p=0$ for different values of the coupling constant $J$ on the first layer.   See Figs.~\ref{fig:rhoc_p0_J},~\ref{fig:rho1_p0_J}.

The spectra show a gap $\Delta$. At $J=0$, we can determine the value of $\Delta$, and the full spectrum, exactly from the solution in Appendix~\ref{app:J0}. At non-zero $J$, the value of $\Delta$ decreases with increasing $J$. We also observe signs of non-analyticities in the spectrum at $\omega = 3 \Delta, 5 \Delta, \ldots$: these are expected at all odd multiples of $\Delta$ from (\ref{sd2}), and are thresholds associated with the decay of an excitation to 3 excitations.

\begin{figure}
\includegraphics[width=3.7in]{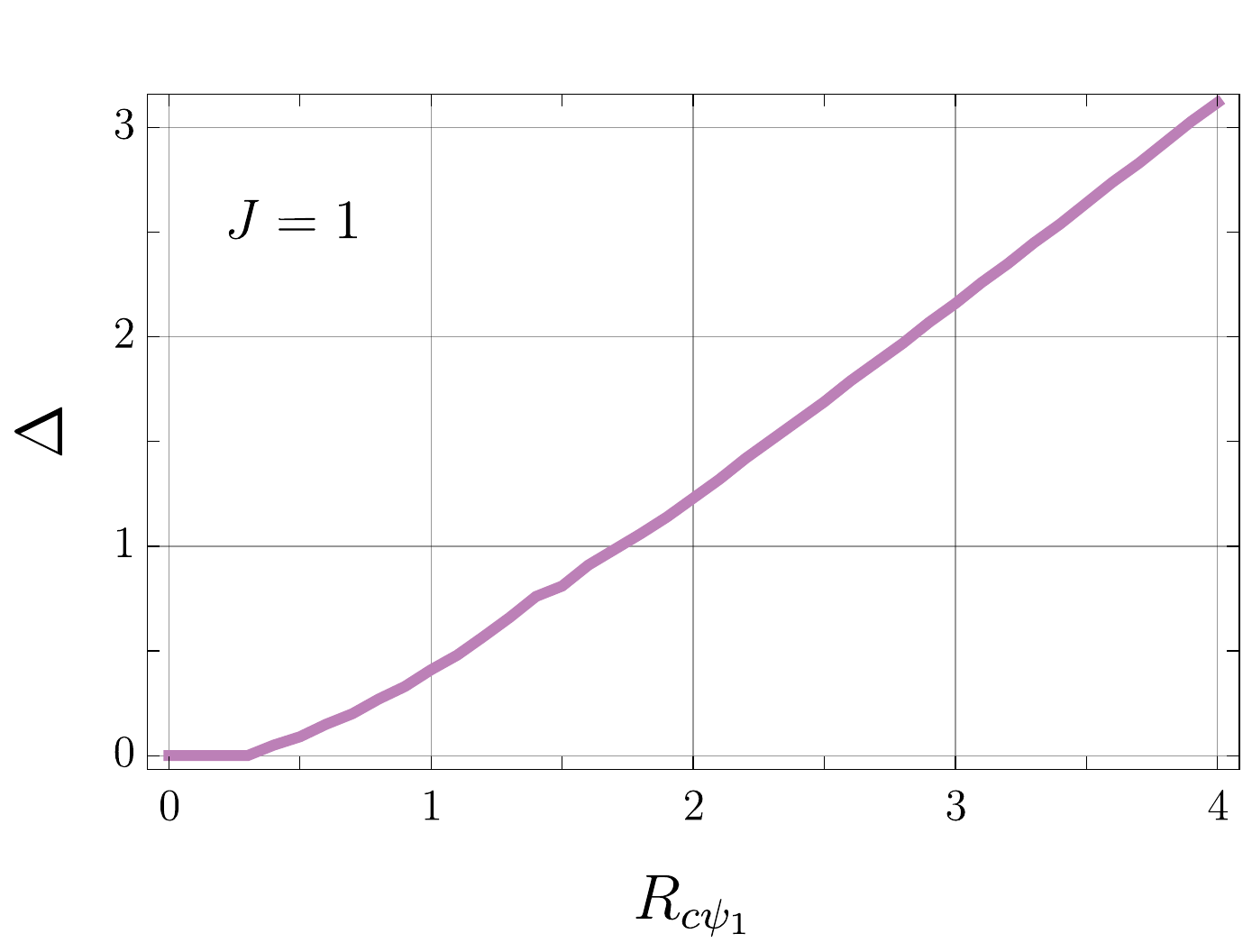}
\caption{Numerically computed gap $\Delta$ as function of $R_{c\psi_1}$ with fixed r.m.s.   exchange in the ancilla layers $J=1$ at doping $p=0$. }
\label{fig:gap_R_J1}
\end{figure}

In conclusion of the analysis at zero doping, we compute the value of the gap $\Delta$ as a function of the off-diagonal self-energy $R_{c\psi}$. In Fig.~\ref{fig:gap_R_J1} we show its behavior at $J=1$.

\subsection{Non-zero doping}
\label{sec:fls_spectral}

\begin{figure}
\centerline{\includegraphics[width=5in]{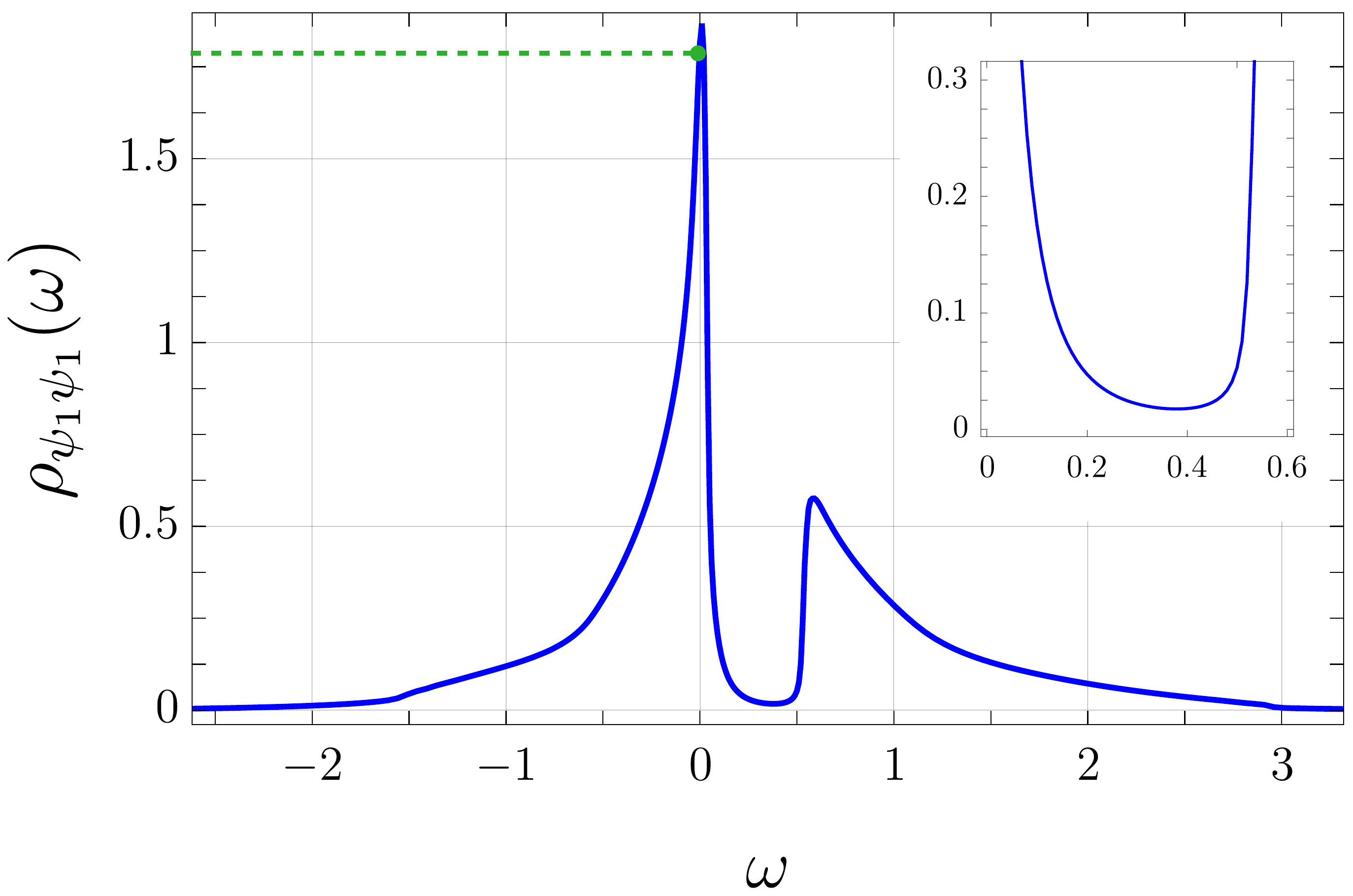}}
\caption{Spectral density of $\Psi_1$ fermion for the case in Fig.~\ref{fig:flsflrhocc}, with $R_{c \psi_1}=0.75$ and doping $p=0.246$. The dashed line indicates the density at $\omega=0$. The r.m.s. exchange in the ancilla layers is $J=1$. Inset: behavior of the spectral density at small frequencies.}
\label{fig:rho1_p25_J1}
\end{figure}

We already showed a result for the spectral function of the FL* phase in Fig.~\ref{fig:flsflrhocc} for $p=0.246$. There is now no strict gap in the spectrum, but a pseudogap at positive frequencies; any non-zero hole doping moves the Fermi energy to the top of the lower band in the insulator in Fig.~\ref{fig:rho1_p0_J}. The density of states at the Fermi energy, $\omega=0$, is suppressed to a value that is constrained by the extended Luttinger theorem in Appendix~\ref{app:luttinger}: as illustrated in Fig.~\ref{fig:flsflrhocc}, and from (\ref{DoSEF}), the density of states at the Fermi level has the same value as the case where $p$ holes are doped in a fully-filled band of the $C$ electrons. For the case with a non-random dispersion $\varepsilon_{\bf k}$, the equivalent statement would be that the Fermi surface encloses a volume equivalent to $p$ holes in the FL* phase. We also show the spectral density of the $\psi_1$ fermions in the first ancilla layers in Fig.~\ref{fig:rho1_p25_J1}. 

\begin{figure}
\centerline{\includegraphics[width=5in]{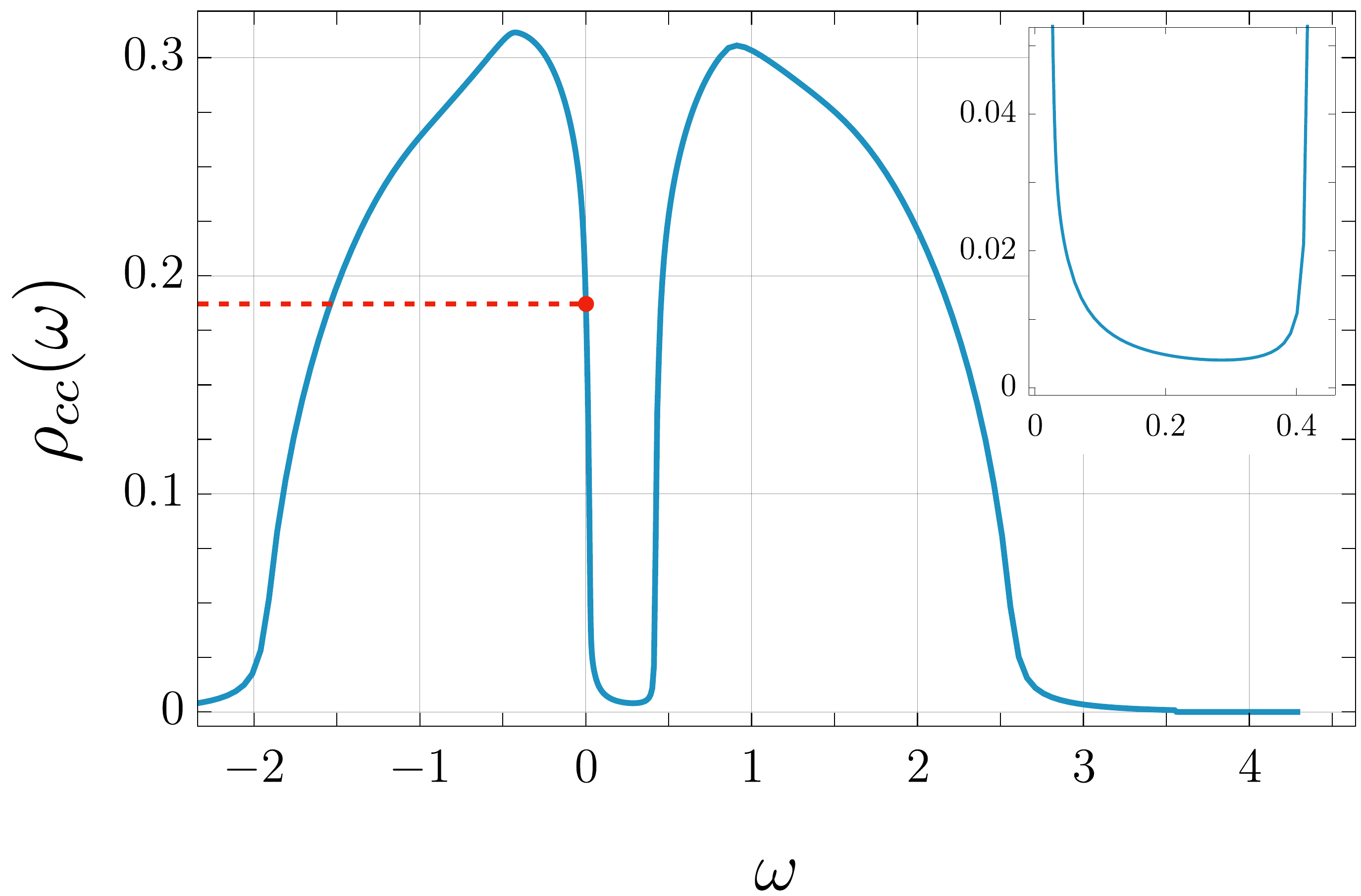}}
\caption{Electron $C$ spectral density at $R_{c \psi_1}=0.75$ and value of doping $p=0.1$. The dashed line indicates the density at $\omega=0$. The r.m.s. exchange in the ancilla layers is $J=1.8$.}
\label{fig:rhoc_p01_J}
\end{figure}
\begin{figure}
\centerline{\includegraphics[width=5in]{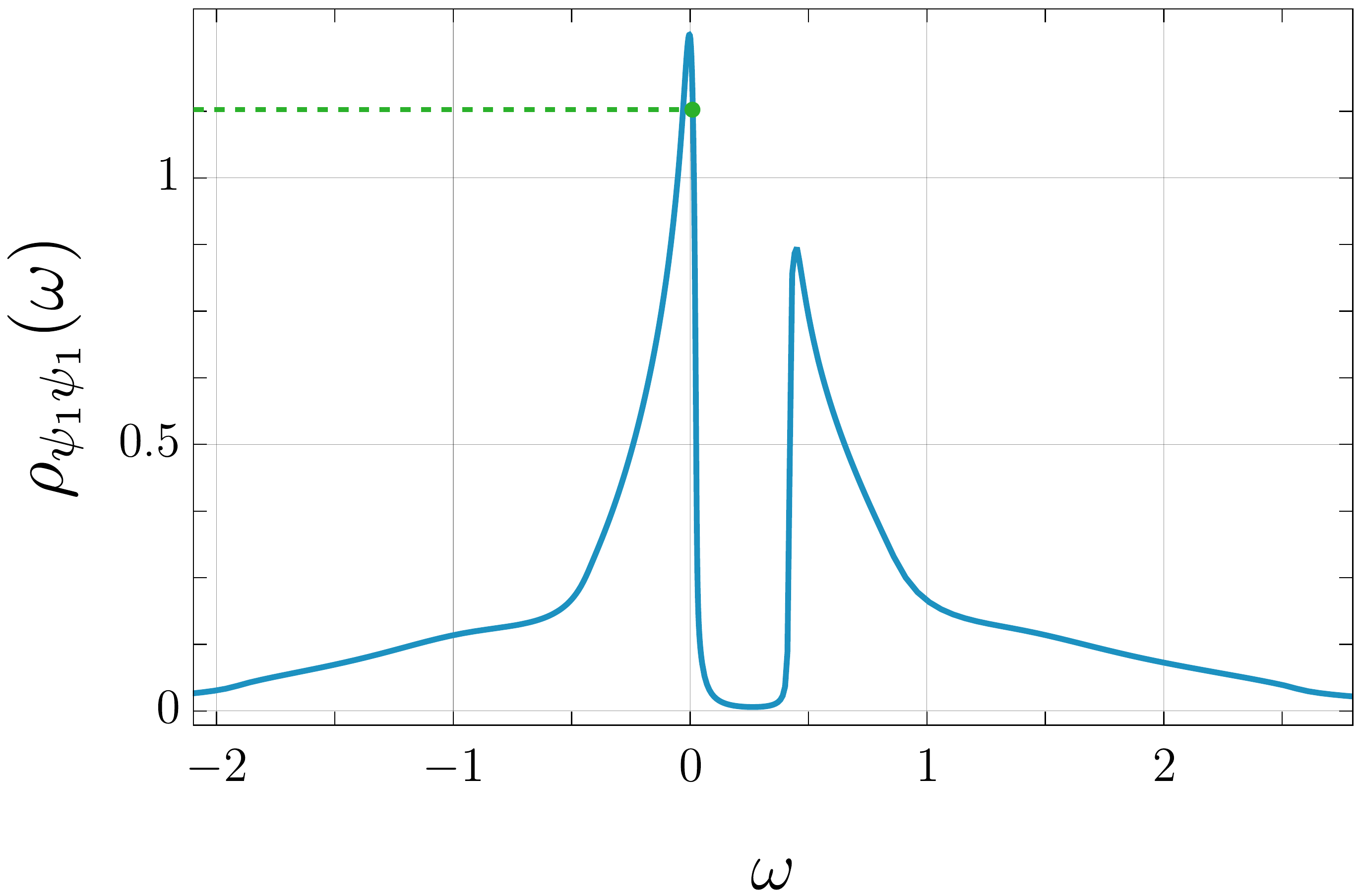}}
\caption{$\Psi_1$ fermion spectral density for the case in Fig.~\ref{fig:rhoc_p01_J}. The dashed line indicates the density at $\omega=0$.}
\label{fig:rho1_p01_J}
\end{figure}
Spectral functions at a smaller doping $p=0.1$ appear in Figs.~\ref{fig:rhoc_p01_J} and \ref{fig:rho1_p01_J}. Note the decrease in the electron density of states at the Fermi level, and pseudogap above the Fermi energy.

We note that the above computations of the FL* spectral functions do not include the influence of the spin liquid state in the second ancilla layer. Such a coupling appears at higher orders in $1/M$, and consequences are similar to those described by Burdin {\it et al.\/} \cite{Burdin_2002} for the Kondo lattice: for the case of a SYK spin liquid on the second ancilla layer, there are marginal Fermi liquid self energy corrections for the quasiparticles on the small Fermi surface.

\section{Conclusions}
\label{sec:conc}

We examined a solvable single band model with hole doping $p$ away from half-filling. The model displays a small Fermi surface of holes of volume $p$ at small doping, and a transition to a large Fermi surface of holes of volume $1+p$ (or equivalently, a Fermi surface of electrons of volume $1-p$) obeying the conventional Luttinger theorem at large doping. This basic phenomenology tracks the physics of the hole doped cuprates in the crossover from the pseudogap at low doping to the Fermi liquid at high doping, as displayed in numerous experiments \cite{CPLT18,Shen19,Ramshaw20}. There are no broken symmetries in any of the phases we find, and so the observed broken symmetries at low temperatures are presumed to be secondary phenomena. Our approach maps the low energy electronic excitations of the pseudogap phase to those a doped Kondo insulator (plus a spin liquid in the second ancilla layer). The popular approach of a doped Mott insulator \cite{LeeWen06} requires a non-perturbative binding of spinons and holons in the FL* state, and that is not needed in our framework. 

We employed an ancilla approach, in which the physical layer is coupled to two fictitious layers of ancilla spins. We show in Appendix~\ref{app:hubbard} that, in a suitable limit, the ancilla spins can be eliminated by a canonical transformation, and the resulting effective Hamiltonian for the physical spins is a single band Hubbard model in the strong correlation regime. So we can view the ancilla spins as being akin to Hubbard-Stratonovich fields, which are chosen to be a pair of quantum spins rather than bosonic fields. The pairing of ancilla spins is essential to avoid introducing new anomalies \cite{Else2020} associated with extended Luttinger theorems. Fluctuations of a SU(2)$_S$ gauge field, acting as a rotating reference frame in spin space \cite{Sachdev:2009cp,MSSS18}, are also need to ensure that the final theory acts only on the physical layer, and the ancilla spins are projected onto rung singlets \cite{Yahui20a,Yahui20b}. 

In the absence of symmetry breaking, the insulator at half-filling ($p=0$) in a single band model is neccesarily a spin liquid with topological order. Our ancilla approach captures {\it both} spin and charge fluctuations in such a Mott insulator in an interesting manner.
The physical electron ($C$) layer and the first ancilla ($\Psi_1$) layer form a Kondo insulator. There is a charge gap in this Kondo insulator, and conventional electron and hole excitations across the charge gap, with no spin-charge separation. The spectrum of these charge excitations is shown in Fig.~\ref{fig:rhoc_p0_J} for the case of random matrix hopping in the electron layer. At the same time, this insulator also has fractionalized spinon excitations---indeed such fractionalized excitations are required by the extended Luttinger theorems. In our approach, these fractionalized excitations reside on the second ancilla ($\Psi_2$) layer. In the present paper we used a SYK$_4$ model of a gapless spin liquid, but other possibilities have also been considered \cite{Yahui20a,Yahui20b}.

Upon doping this Mott insulator, we obtain a FL* phase as our theory for the pseudogap. Given the mapping of the Mott insulator to the Kondo insulator above, the FL* phase maps onto a doped Kondo insulator. The spectrum of electron-like excitations of this metallic phase are shown in Figs.~\ref{fig:flsflrhocc} and \ref{fig:rhoc_p01_J} for two values of $p$. These are computed for the simplest case where the band structure of the physical electron layer is a random matrix---so the density of states in the large doping Fermi liquid phase will be a Wigner semi-circle, as shown in Fig.~\ref{fig:flsflrhocc}. In the FL* phase, our results show a number of notable features: a reduction in the density of states at the Fermi level, a pseudogap above the Fermi level, and a pronounced particle-hole asymmetry. It would be interesting to extend these computations to more realistic band structures on the physical layer; computations with a realistic band structure were carried out in Ref.~\cite{Yahui20a}, but without the dynamic spin fluctuations present in our SYK model. 

In comparing to experimental observations of pseudogap spectra in STM experiments \cite{Seamus04,LeeDavis09,Seamus12}, we do observe a particle-hole asymmetry with the same sign. Moreover, as in Figs.~\ref{fig:flsflrhocc} and \ref{fig:rhoc_p01_J}, the minimum in the local density of states (LDOS) is indeed observed to be slightly above the Fermi level at higher temperatures (see Fig.~3C in Lee {\it et al.} \cite{LeeDavis09}). As the temperature is lowered, the minimum in the LDOS moves towards the Fermi level, indicating the appearance of physics not captured by our present analysis. It would be interesting to study fluctuation corrections, possibly from spinon or electron pairing, or from disorder \cite{LRRMP}, and determine if they can explain the pinning of the LDOS minimum to the Fermi level as $T \rightarrow 0$.

At larger $p$ our model undergoes a first order phase transition to a conventional Fermi liquid phase (FL), as we showed in Section~\ref{sec:numerical}. In this phase, the physical electronic layer is largely decoupled from the two ancilla layers, which are locked into a trivial spin gap insulator, as illustrated in Fig.~\ref{fig:layers}b. 
The first order transition is compatible with the observed sudden change between incoherent and coherent photoemission spectra in the antinodal region upon a small change in doping in Bi2122 \cite{Shen19}; we also that the sharp vertical boundary of the pseudogap phase in the doping-temperature plane, as detected by the nematicity in X-ray scattering \cite{Hawthorn20}.
On the other hand, the critical fluctuation effects associated with ghost Fermi surfaces, studied in Refs.~\cite{Yahui20a,Yahui20b} are absent in the present large $M$ limit at $T=0$. It is possible that such fluctuations are restored at non-zero temperatures, and it would be interesting to incorporate such fluctuations in extensions of our approach.

Our computations also make general predictions for spectra observed in photoemission and neutron scattering experiments in the pseudogap phase. While, the ancillas are a computational device, they also give a simple physical picture for such experiments:
\begin{itemize}
\item 
For photoemission, the main prediction is that the electronic spectrum near the Fermi surface should be similar to that of a lightly doped Kondo insulator. 
This provides a direct understanding of the spectrum, rather than proceeding by doping the spin liquid of a Mott insulator \cite{LeeWen06}.
\item For neutron scattering, the prediction is that there are 2 components to the spin fluctuations. One component consists of the spin fluctuations of the particles/holes observed in photoemission, which would also be present in a doped Kondo insulator. The other component arises from the spinons of the spin liquid in the second ancilla layer, and this is not present in a doped Kondo insulator.
\end{itemize}

In closing, we note an interesting correspondence along the lines of Ref.~\cite{SS10}, to a recent study by Sahoo {\it et al.} \cite{Sahoo:2020unu} of wormholes and Hawking-Page transitions in coupled SYK models. They considered two $q=4$ complex SYK models with random 4-fermion interactions determined by the same couplings, and a non-random 2-fermion coupling between them. Under suitable conditions, they found a first-order transition between two compressible non-Fermi liquid phases. In the `large black hole' phase, all fermions are involved in the low energy non-Fermi liquid excitations and a wormhole connects to the black holes dual to the SYK models. This is separated by a first order {\it partial\/} Hawking-Page transition
from a `small black hole' phase in which a particular linear combination of fermions is locked into a trivial gapped state, while the remaining fermions form the non-Fermi liquid. This has parallels in our study, although the details are different. We have two $q=4$ SYK models with different random 4-fermion couplings, one $q=2$ SYK model with random 2-fermion couplings, and non-random 4-fermion couplings between the SYK models. The FL phase is the analog of the small black hole phase: in our case, the $q=4$ SYK models lock into a trivial insulator, while the $q=2$ SYK model forms a Fermi liquid state. The FL* phase is the analog of the large black hole phase:
in our case, the $q=2$ SYK model and one $q=4$ SYK model together form a Fermi liquid, while the other $q=4$ SYK model forms a non-Fermi liquid. 

For a closer holographic analogy, we can imagine 3 SYK models in a row, with the central SYK model coupled to the outer ones. In one phase, the central black hole is connected by a wormhole to the one on the left, and in the other phase the central black hole is connected by a wormhole to the one on the right. These phases are separated by a Hawking-Page type transition, which is the holographic analog of the FL* to FL transition discussed here.

\subsection*{Acknowledgement}
\label{sec:ack}

We are grateful to Sudi Chen, Seamus Davis, and Z.-X. Shen for enlightening discussions on the spectrum of the pseudogap phase in the cuprates, and Sharmistha Sahoo for pointing out the connection to Ref.~\cite{Sahoo:2020unu}. We also thank Sasha Brownsberger, Piers Coleman, Ilya Esterlis, Antoine Georges, Grigory Tarnopolsky, and Pengfei Zhang for valuable discussions. 
This research was supported by the National Science Foundation under Grant No.~DMR-2002850. 
This work was also supported by the Simons Collaboration on Ultra-Quantum Matter, which is a grant from the Simons Foundation (651440, S.S.).



\appendix

\section{Mapping to the single band Hubbard model}
\label{app:hubbard}

In this appendix we obtain the effective Hamiltonian for non-interacting electrons $C$ in the physical layer coupled to two ancilla layers, as in Fig.~\ref{fig:layers}, in the limit of large $J_{\perp}$, for the SU(2) case with $M=2$. To leading order in the $1/J_{\perp}$ expansion, this effective Hamiltonian turns out to be the familiar Hubbard model. This Hubbard model can be in a strong correlation regime by a judicious choice of ancilla couplings, as we shall show below. 

For simplicity, we only consider the case with non-random, nearest-neighbor, exchange interactions. So we have
\begin{itemize}
\item non-interacting electrons at chemical potential $\mu$ with nearest-neighbor hopping $t$ in the physical layer,
\item antiferromagnetic exchange $J_K$ between the physical layer and the first ancilla layer,
\item antiferromagnetic exchange $J_\perp$ between the second ancilla layer and the first ancilla layer,
\item antiferromagnetic exchange $J_1$ within the first ancilla layer,
\item antiferromagnetic exchange $J_2$ within the second ancilla layer.
\end{itemize}

We can perform Schrieffer-Wolff transformation \cite{Bravyi11} in powers of $1/J_\perp$ to eliminate the ancilla layers. To order $1/J_\perp^2$, this will yield an effective Hamiltonian for the physical layer of $C$ fermions of the following form
\bea
H_{\rm eff} &=&  \sum_i \left[E_0 -\mu_{\rm eff} \, C_{i;\alpha}^\dagger C_{i; \alpha} + U_{\rm eff}\, C_{i; \uparrow}^\dagger C_{i;\uparrow} C_{i; \downarrow}^\dagger C_{i;\downarrow}\right] \nonumber \\
&~&~+ \sum_{\langle ij\rangle}
\left[
-t \left(C_{i; \alpha}^\dagger C_{j;\alpha} + \mbox{H.c.} \right)
+\frac{J_{\rm eff}}{4} \, C_{i;\alpha}^\dagger \vec{\sigma}_{\alpha\beta} C_{i; \beta}\, C_{j;\gamma}^\dagger \vec{\sigma}_{\gamma\delta} C_{j; \delta} 
\right]
\,,
\eea
where $\vec{\sigma}$ are the Pauli matrices.

At order $1/J_{\perp}$, we only introduce on-site couplings in the physical layer. These couplings can be computed by exact diagonalization of the 3 site model, with one site in each layer. With 0 or 2 electrons in the $C$ layer, the ancilla spins lock in a singlet with energy $-3 J_\perp/4$. With 1 electron in the physical layer, the ground state energy of the 3 site model is 
\beq
E_3 = -\frac{J_\perp + J_K}{4} - \frac{1}{2} \left(
J_\perp^2 + J_K^2 - J_\perp J_K \right)^{1/2}\,.
\eeq
This lowers the energy of a singly-occupied site in the physical layer, and the result is an effective {\it repulsive} interaction between the electrons; by matching energy levels to those of $H_{\rm eff}$, we obtain
\bea
\mu_{\rm eff} &=& \mu + U_{\rm eff}/2 \nonumber \\
U_{\rm eff} &=& \frac{3J_K^2}{8 J_\perp} + \frac{3J_K^3}{16 J_\perp^2} + \mathcal{O}(1/J_\perp^3) \,. 
\eea

The exchange coupling $J_{\rm eff}$ appears only at order $1/J_\perp^2$, and it can be computed by diagonalizing a 6 site cluster with 2 sites in each layer. We performed such a diagonalization in a power series in $1/J_\perp$ and obtained
\bea
E_0 &=& - \frac{3 J_\perp}{2} - \frac{3 (J_1 + J_2)^2}{64 J_\perp} - \frac{3 (J_1 + J_2)^3}{256 J_\perp^2}+ \mathcal{O}(1/J_\perp^3) \nonumber \\
J_{\rm eff} &=& \frac{J_K^2 (J_1 + J_2)}{4J_\perp^2}+ \mathcal{O}(1/J_\perp^3)\,.
\eea

We now observe that we can obtain the physically reasonable heierarchy $J_{\rm eff} \ll t \ll U_{\rm eff}$ by choosing
\beq
\left(\frac{J}{J_\perp} \right)^3 \ll \frac{t}{J_\perp} \ll \left(\frac{J}{J_\perp} \right)^2 \ll 1 \,,
\eeq
where $J$ is a generic coupling of order $J_K$, $J_1$, or $J_2$.

\section{Derivation of Schwinger-Dyson equations}
\label{app:saddle}

After an averaging of the initial Hamiltonian (\ref{eq:initial_Hamiltonian}) over the random couplings we obtain the action:
\begin{align}
S=S_B + S_t+S_{J}+S_{J_K}
\end{align}
where the kinematic Berry phase term is 
\begin{align}
 S_{B}=\sum_{i,\alpha}\int d\tau C^{\dagger}_{i;\alpha}(\tau)(\partial_\tau-\mu_c) C_{i ;\alpha}(\tau)+\sum_{i,a,\alpha}\int d\tau \Psi^\dagger_{i;a;\alpha}(\tau)(\partial_\tau -\mu_{\psi_a})\Psi_{i;a;\alpha}(\tau)\,,
\end{align}
the random hopping term is 
\begin{align}
S_{t}=-\sum_{i,j}^{N} \frac{t^2}{2N}\int d\tau d\tau' C^\dagger_{i;\alpha}(\tau) C_{j;\alpha}(\tau )C^\dagger_{j;\beta}(\tau')C_{i;\beta}(\tau') \,,
\end{align}
the random exchange terms are
\begin{align}
S_{J}=-\sum_{a,i\neq j}^{N} \frac{J^2}{4 N M}\int d\tau d \tau' S_{i;a;\alpha \beta}(\tau) S_{j;a;\beta \alpha}(\tau) S_{i;a; \alpha' \beta'}(\tau') S_{j;a;\beta' \alpha'}(\tau') \,,
\end{align}
and the non-random exchange terms are
\begin{align}
S_{J_K}=\frac{J_\perp}{M} \sum_i  \int d\tau S_{i;1; \alpha \beta}(\tau) S_{i;2;\beta \alpha}(\tau)+\frac{J_K}{M} \sum_i \int d\tau C^\dagger_{i;\alpha}(\tau) C_{i;\beta}(\tau) S_{i;1;\beta \alpha}(\tau)
\end{align}

The random hopping term can be rewritten in terms of a bilocal field:
\begin{align}
 G^{\alpha \beta}_{cc}(\tau,\tau')=-\frac{1}{N}\sum_i C_{i;\alpha}(\tau)C^{\dagger}_{i;\beta}(\tau')\,.
\end{align}
Then, assuming $G^{\alpha \beta}_{cc}=G_{cc} \delta_{\alpha,\beta}$:
\begin{equation}
S_{t}/(N M)=\int d\tau d\tau' \left[\frac{t^2}{2} G_{cc}(\tau,\tau') G_{cc}(\tau',\tau)-\Sigma_{cc}(\tau',\tau)\left(G_{cc}(\tau,\tau')+\frac{1}{N M}\sum_{i,\alpha} C_{i;\alpha}(\tau)C^\dagger_{i;\alpha}(\tau')\right)\right]
\end{equation}

To simplify the random exchange terms, we introduce the following 4-field:
\begin{equation}
 Q_a^{\alpha \beta \alpha' \beta'}(\tau,\tau')=-\frac{1}{N}\sum_i S_{i;a;\alpha\beta}(\tau)S_{i;a;\alpha'\beta'}(\tau)(\tau')\,.
\end{equation}
Including a corresponding 4-self-energy we obtain:
\begin{align}
S_{J}= & \int d\tau d\tau' \left[-\frac{N J^2}{4  M} Q_a^{\alpha \beta \alpha'\beta'}(\tau,\tau')Q_a^{ \beta \alpha \beta' \alpha' }(\tau,\tau')  \right. \notag \\
&~~~~~~\left. 
-\frac{N}{M}\Sigma_{Q;a}^{\alpha \beta \alpha' \beta'}(\tau',\tau)\left( Q_a(\tau,\tau')^{\alpha \beta \alpha' \beta'}+\frac{1}{N}\sum_i S_{i;a;\alpha\beta}(\tau) S_{i;a;\alpha'\beta'}(\tau')\right)\right]\,.
\end{align}
In the large $M$ limit, we can safely assume that the saddle-point has \cite{sy} $Q_a^{\alpha \beta \alpha'\beta'}=\delta_{\alpha\beta'}\delta_{\beta\alpha'}Q_a$, and similarly for the corresponding self energy.
We also introduce a bilocal field:
\begin{equation}
 G^{i}_{\psi_a \psi_a}(\tau,\tau')=-\frac{1}{M}\sum_{\alpha} \Psi_{i;a;\alpha}(\tau)\Psi^\dagger_{i;a;\alpha}(\tau')\,.
\end{equation}
Then we obtain,
\begin{equation}
\begin{split}
S_{J}/(N M)= \int d\tau d\tau' \Bigg[-\frac{ J^2}{4} Q_a(\tau,\tau')^2
-\Sigma_{Q;a}(\tau',\tau)\left( Q_a(\tau,\tau')- G_{\psi_a \psi_a}(\tau,\tau')G_{\psi_a \psi_a}(\tau',\tau) \right)&\\
-\Sigma_{\psi_a \psi_a}(\tau',\tau)\left(G_{\psi_a \psi_a}(\tau,\tau')+\frac{1}{N M}\sum_{i,\alpha} \Psi_{i;a;\alpha}(\tau)\Psi^\dagger_{i;a;\alpha}(\tau')\right)\Bigg]\,. &
\end{split}
\end{equation}

The non-random exchange terms can be rewritten in terms of bilocal fields and self-energies as follows.
We introduce Green's functions as:
\begin{equation}
  G^{i}_{c \psi_a}(\tau,\tau')=-\frac{1}{M}\sum_\alpha C_{i; \alpha}(\tau)\Psi^{\dagger}_{i;a;\alpha}(\tau') \quad \quad G^{i}_{\psi_a c}(\tau,\tau')=-\frac{1}{M}\sum_\alpha \Psi_{i;a;\alpha}(\tau)C^\dagger_{i;\alpha}(\tau')
\end{equation}
We assume that $G_{c\psi_a}(\tau=+0)=G_{c\psi_a}(\tau=-0)$ due to commutation relation. Indeed $G_{c\psi_a}(\tau)=-\langle T C(\tau)\Psi_a^{\dagger}(0)\rangle$. Then $G_{c\psi_a}(+0)=-\langle C \Psi_a^{\dagger}\rangle$ and $G_{c\psi_a}(-0)=\langle  \Psi_a^{\dagger} C\rangle=-\langle C \Psi_a^{\dagger} \rangle$.
Then, after assuming $G^i=G$ we obtain
\begin{equation}
\begin{split}
&S_{J_K}/(NM)= \int d\tau d\tau' \Bigg[ -J_K G_{\psi_1 c}(\tau+0,\tau) G_{c \psi_1}(\tau+0,\tau)\delta(\tau-\tau') \\
&\qquad \qquad \qquad -\Sigma_{\psi_1 c}(\tau',\tau)\left(G_{\psi_1  c}(\tau,\tau')+\frac{1}{N M}\sum_{i,\alpha} \Psi_{i;1;\alpha}(\tau)C^\dagger_{i;\alpha}(\tau')\right) \\
&\qquad \qquad \qquad -\Sigma_{c \psi_1}(\tau',\tau)\left(G_{c \psi_1}(\tau,\tau')+\frac{1}{N M}\sum_{i,\alpha} C_{i;\alpha}(\tau) \Psi^\dagger_{i;\alpha}(\tau')\right) \Bigg]\\
&\qquad \qquad \qquad + \int d\tau d\tau' \Bigg[ -J_\bot G_{\psi_2\psi_1}(\tau+0,\tau) G_{\psi_1\psi_2}(\tau+0,\tau)\delta(\tau-\tau') \\
&\qquad \qquad \qquad-\Sigma_{\psi_2\psi_1}(\tau',\tau)\left(G_{\psi_2\psi_1}(\tau,\tau')+\frac{1}{N M}\sum_{i,\alpha} \Psi_{i;2;\alpha}(\tau)\Psi^{\dagger}_{i;1;\alpha}(\tau')\right) \\
&\qquad \qquad \qquad
-\Sigma_{\psi_1\psi_2}(\tau',\tau)\left(G_{\psi_1\psi_2}(\tau,\tau')+\frac{1}{N M}\sum_{i,\alpha} \Psi_{i;1;\alpha}(\tau)\Psi^\dagger_{i;2;\alpha}(\tau')\right) \Bigg] \\
\end{split}
\end{equation}
Now we can integrate over $\tilde{\Psi}=(C,\Psi)$ degrees of freedom. The action is quadratic in these variables: $S=\sum_{i,\alpha,n} 
\tilde{\Psi}^\dagger_{i,\alpha,n} H_{i,\alpha,n} \tilde{\Psi}_{i,\alpha,n}$, where the $n$ subscript refers to Matsubara frequency:
\begin{equation}
 H_n =
\left(
\begin{array}{ccc}
-i\omega_n-\mu+\Sigma_{cc,n}  & \Sigma_{c\psi_1,n} & 0\\
\Sigma_{\psi_1 c,n} & -i\omega_n-\mu_{\psi_1}+\Sigma_{\psi_1\psi_1  ,n} &  \Sigma_{\psi_1\psi_2,n}\\
0  &  \Sigma_{\psi_2\psi_1 ,n} &  -i\omega_n-\mu_{\psi_2}+\Sigma_{\psi_2 \psi_2 ,n}
\end{array}
\right)
\end{equation}
After integrating these degrees of freedom we obtain:
\begin{equation}
 S/(N M)=\sum_n \left( -\log(\text{det}H_n)-\Sigma_{i j,n}G_{ij,n}\right) + \frac{1}{2} \left(\mu_{\psi_1} + \mu_{\psi_2} \right)\,.
\end{equation}
Differentiating over the self-energies we obtain:
\begin{equation}
 G_n=-
\left(
\begin{array}{ccc}
-i\omega_n-\mu+\Sigma_{cc,n}  & \Sigma_{c\psi_1,n} & 0\\
\Sigma_{\psi_1 c,n} & -i\omega_n-\mu_{\psi_1}+\Sigma_{\psi_1\psi_1  ,n} &  \Sigma_{\psi_1\psi_2,n}\\
0  &  \Sigma_{\psi_2\psi_1 ,n} &  -i\omega_n-\mu_{\psi_2}+\Sigma_{\psi_2 \psi_2 ,n}
\end{array}
\right)^{-1}\,.
\label{green_funct}
\end{equation}
Differentiating over the Green's functions we obtain the following self-energies:
\begin{equation}
 \begin{split}
 &\Sigma_{cc}(\tau)= t^2 G_{cc}(\tau)\\
 &\Sigma_{\psi_a \psi_a}(\tau)=- J^2G_{\psi_a\psi_a}(\tau)^2G_{\psi_a \psi_a}(-\tau)\\
 &\Sigma_{c \psi_1}(\tau)=-J_K G_{\psi_1 c}(\tau=+0)\delta(\tau)\\
 &\Sigma_{\psi_1c }(\tau)=-J_K G_{c \psi_1}(\tau=+0)\delta(\tau)\\ 
  &\Sigma_{\psi_1 \psi_2}(\tau)=-J_{\bot}G_{\psi_2 \psi_1}(\tau=+0)\delta(\tau)\\
 &\Sigma_{\psi_2 \psi_1}(\tau)=-J_{\bot}G_{\psi_1\psi_2}(\tau=+0)\delta(\tau)
 \end{split}
 \label{selfenergy}
\end{equation}
which leads to the expressions in Section~\ref{sec:sd}.

\subsection{Free energy}
The free energy is given by the action at the saddle point: $\beta F=S_{\text{saddle}}$
\begin{align}
\begin{split}
&\frac{\beta F}{N M}=\sum_n \left( -\log(\text{det}H_n)-\Sigma_{i j,n}G_{ij,n}\right)\\
&+\int d\tau d\tau' \left[\frac{t^2}{2} G_{cc}(\tau,\tau') G_{cc}(\tau',\tau )-\frac{ J^2}{4}G^2_{\psi_a \psi_a}(\tau,\tau')G^2_{\psi_a \psi_a}(\tau',\tau) 
\right]\\
&-\beta J_K G_{\psi_1 c}(0) G_{c \psi_1}(0)-\beta J_\perp   G_{\psi_2\psi_1}(0) G_{\psi_1\psi_2}(0)+\frac{1}{2} \left(\mu_{\psi_1} + \mu_{\psi_2} \right)\\
\end{split}
\end{align}
The double integration can be further simplified assuming that Green's functions depend on time difference(I also divide it by $i \omega_n$ to get rid of divergences):
\begin{align}
\begin{split}
&\frac{\beta F}{N M}=\sum_n \left( -\log(\text{det}H_n/(i\omega_n)^3)-\Sigma_{i j,n}G_{ij,n}\right)\\
&+\int_0^\beta d\tau  (\beta-\tau)\left[t^2G_{cc}(\tau) G_{cc}(-\tau)-\frac{J^2}{2}G^2_{\psi_a \psi_a}(\tau)G^2_{\psi_a \psi_a}(-\tau) 
\right]\\
&-\beta J_K G_{\psi_1 c}(0) G_{c \psi_1}(0)-\beta J_\perp   G_{\psi_2\psi_1}(0) G_{\psi_1\psi_2}(0)+\frac{1}{2} \left(\mu_{\psi_1} + \mu_{\psi_2} \right)\\
\end{split}
\end{align}
The formula can be checked (at least partially) in the following way: we compute Free energy for different $\Sigma_{i j}$ and it the minimum of the Free energy should coincide with the actual solution of the Schwinger-Dyson equations.

\section{Luttinger relations}
\label{app:luttinger}

This appendix will employ a conventional Luttinger-Ward formalism to obtain the distinct Luttinger relations in the FL* and FL phases in a unified manner. All results here are exact, and hold to all orders in the $1/M$ expansion.

First, we review the derivation of the Luttinger result (\ref{muval}) from Refs.~\cite{hewson_book,PG98,Burdin_2000} in the context of our FL* phase where $R_{c\psi_1} \neq 0$, but $R_{\psi_1 \psi_2} = 0$.
We use (\ref{sd10}, \ref{sd11}) to write the $c$ fermion 
Green's function as
\beq
G_{cc}^F (\omega) = \frac{d}{d \omega}  \int_{-\infty}^{\infty} d \Omega D(\Omega) \ln \left[\omega + \mu - R_{c \psi_1}^2 \mathcal{G}_{\psi_1}^F (\omega) - \Omega \right] + R_{c \psi_1}^2 \frac{d \mathcal{G}_{\psi_1}^F (\omega)}{d \omega} G_{cc}^F (\omega), \label{lt1}
\eeq
where the superscript $F$ denotes Feynman Green's functions at $T=0$.
From (\ref{sd8}) we have
\beq
\frac{d \mathcal{G}_{\psi_1}^F (\omega)}{d \omega} = - \left[\mathcal{G}_{\psi_1}^F (\omega) \right]^2 \left(1 - \frac{d \Sigma_{\psi_1 \psi_1}^F (\omega)}{d \omega} \right) \label{lt2}
\eeq
Combining (\ref{lt1}, \ref{lt2}) with (\ref{sd8}, \ref{sd10a}), we obtain for the sum of the $c$ and $\psi_1$ Green's functions
\bea
G_{cc}^F (\omega) + G_{\psi_1 \psi_1}^F (\omega) &=& \frac{d}{d \omega}  \int_{-\infty}^{\infty} d \Omega D(\Omega) \ln \left[\omega + \mu - R_{c \psi_1}^2 \mathcal{G}_{\psi_1}^F (\omega) - \Omega \right] \nonumber \\
&~&~~~~+ G_{\psi_1 \psi_1}^F \frac{d \Sigma_{\psi_1 \psi_1}^F (\omega)}{d \omega} + \frac{d}{d\omega} \ln \left[ \omega + \mu_{\psi_1} - \Sigma_{\psi_1 \psi_1}^F (\omega) \right]\,. \label{lt3}
\eea
Recall that we are in the FL* phase where $R_{\psi_1 \psi_2} = 0$.
Now we can compute the total number of fermions by
\beq
\frac{2-p}{2} = \int_{-\infty}^{\infty} \frac{ d \omega}{2 \pi i} \left[ G_{cc}^F (\omega) + G_{\psi_1 \psi_1}^F (\omega) \right] e^{i \omega 0^+} \label{lt4}
\eeq
As in traditional proofs of the Luttinger theorem \cite{AGD}, the central point is that the frequency integral of the $G_{\psi_1 \psi_1}^F ({d \Sigma_{\psi_1 \psi_1}^F (\omega)}/{d \omega})$ term of (\ref{lt3}) vanishes. In our case, this follows directly from (\ref{sd2}) or from the $G$-$\Sigma$ theory in Appendix~\ref{app:saddle}: this shows that $G_{\psi\psi}^F (\omega) = \delta S/\delta \Sigma_{\psi_1 \psi_1}^F (\omega)$, and so the noted term in (\ref{lt3}) is a total derivative of $\omega$. The remaining terms in (\ref{lt3}) are explicitly total derivatives of $\omega$, and so their frequency integrals are easily evaluated \cite{AGD,PG98}. For $0<p<1$, the FL* phase appears in a regime where the frequency integral of the $
\ln \left[\omega + \mu_{\psi_1} \cdots \right]$ term in (\ref{lt3}) vanishes, and then the
$\ln \left[\omega + \mu \cdots \right]$ term in (\ref{lt3}) yields the FL* case of the Luttinger relation in (\ref{EFval}). We note that this Luttinger relation is found to be accurately obeyed in all our numerical analyses.

Next, let us also consider the FL case (C) in Section~\ref{sec:luttinger} where  $R_{c\psi_1} \neq 0$ and $R_{\psi_1 \psi_2} \neq 0$. Then, from the expressions in Section~\ref{sec:luttinger}, the identity (\ref{lt3}) is replaced by
\bea
G_{cc}^F (\omega) + G_{\psi_1 \psi_1}^F (\omega) &+& 
G_{\psi_2 \psi_2}^F (\omega) = \frac{d}{d \omega}  \int_{-\infty}^{\infty} d \Omega D(\Omega) \ln \left[\omega + \mu - R_{c \psi_1}^2 \mathcal{G}_{\psi_1}^F (\omega) - \Omega \right] \nonumber \\
&+& G_{\psi_1 \psi_1}^F \frac{d \Sigma_{\psi_1 \psi_1}^F (\omega)}{d \omega} + \frac{d}{d\omega} \ln \left[ \omega + \mu_{\psi_1} - \Sigma_{\psi_1 \psi_1}^F (\omega) - R_{\psi_1 \psi_2}^2 \mathcal{G}_{\psi_2}^F (\omega) \right] \nonumber \\
&+& G_{\psi_2 \psi_2}^F \frac{d \Sigma_{\psi_2 \psi_2}^F (\omega)}{d \omega} + \frac{d}{d\omega} \ln \left[ \omega + \mu_{\psi_2} - \Sigma_{\psi_2 \psi_2}^F (\omega)  \right] \,. \label{lt5}
\eea
The frequency integral of (\ref{lt5}) can be performed exactly, as for (\ref{lt3}): now the $ \ln \left[ \omega + \mu_{\psi_1}  \cdots \right]$ term yields unity, while the $ \ln \left[ \omega + \mu_{\psi_2}  \cdots \right]$ term yields 0, and we obtain the FL case of the Luttinger relation in (\ref{EFval}).

For our purposes, the FL case in Fig.~\ref{fig:layers} actually corresponds to case (A) in Section~\ref{sec:luttinger} with $R_{c \psi_1} = 0$ and $R_{\psi_1\psi_2} \neq 0$.
In this case the Luttinger relations follow as special cases of the analyses above.
The Luttinger relations for the
physical $C$ layer follows directly from (\ref{lt1}), where the last term vanishes. The Luttinger relation for the two ancilla layers in obtained from (\ref{lt5}) after dropping the first terms from both the left and right hand sides.

\section{Solution at $J=0$}
\label{app:J0}

\begin{figure}
(a)\includegraphics[width=3.2in]{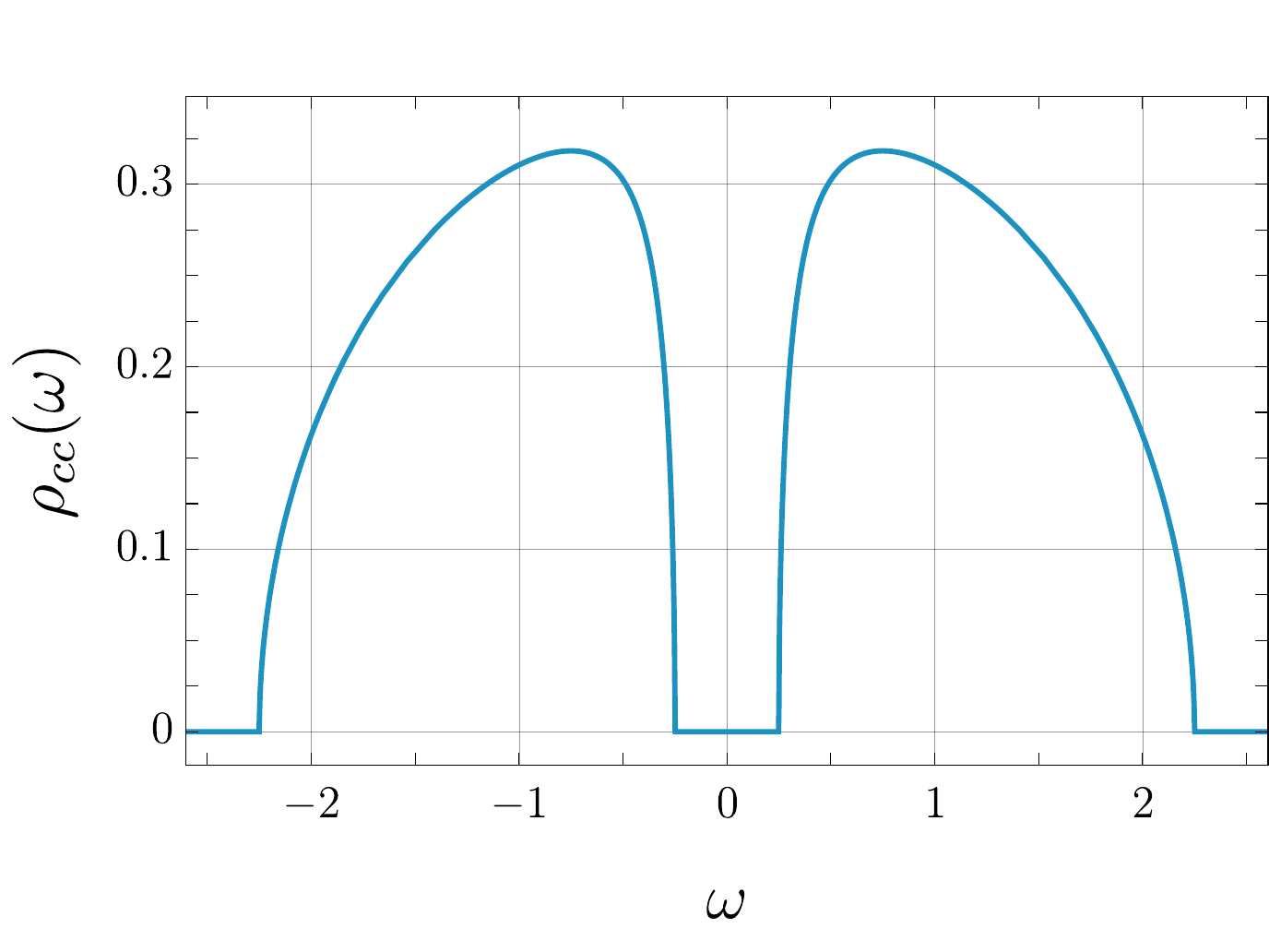}\,\,
(b)\includegraphics[width=3.2in]{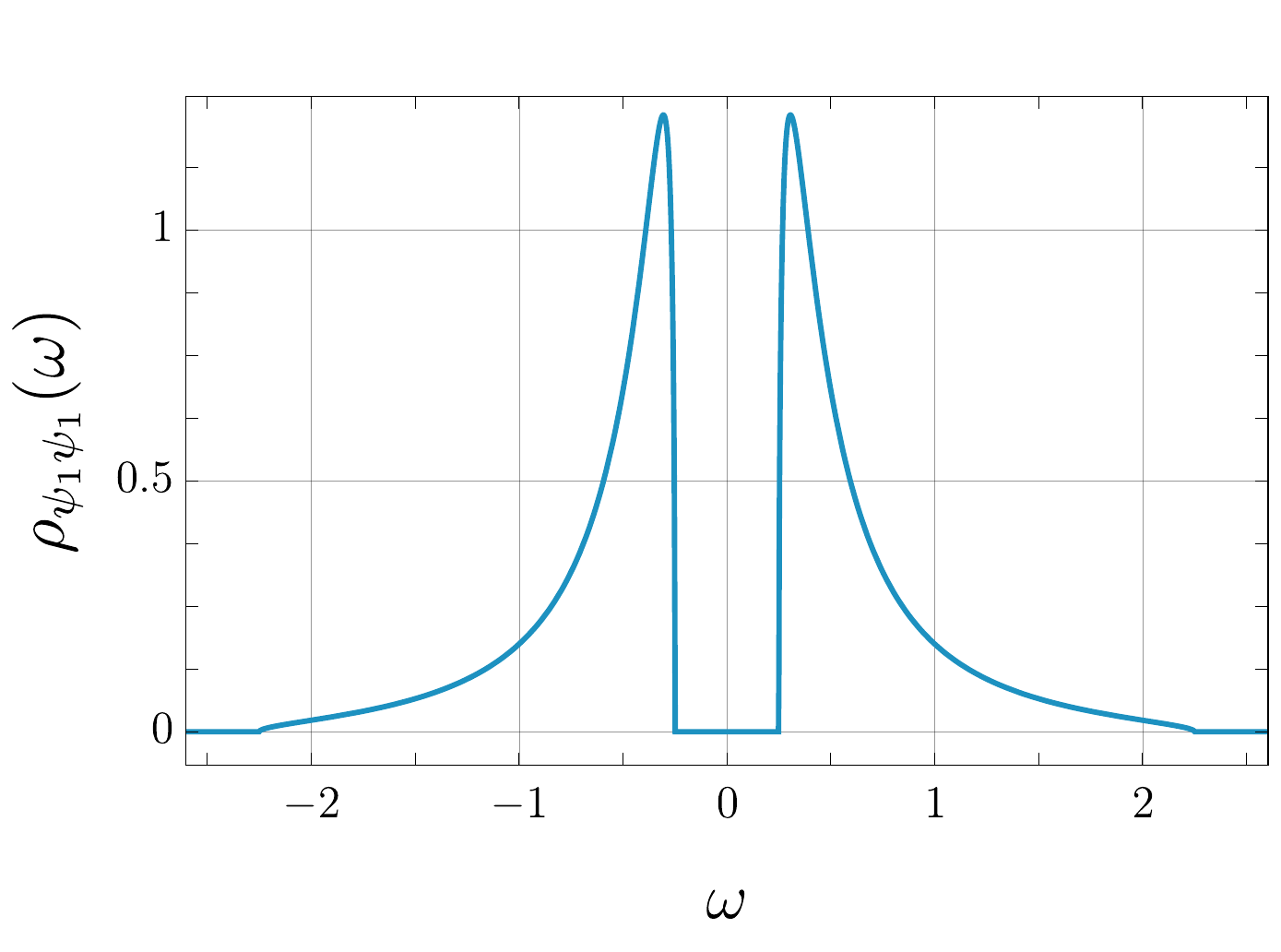}
\caption{Electron (a) and fermionic (b) spectral densities of the insulating solution at $p=0$ for $J=0$, $t=1$, $R_{c \psi_1} = 0.75$.}
\label{fig:ex1}
\end{figure}

It is instructive to examine the solution of the saddle-point equations in the limit where the ancilla spins are decoupled from each other. An exact solution is possible, similar to the Kondo model studied in Ref.~\cite{Burdin_2000}, but now for the case of a Wigner semi-circle band of conduction electrons. The exact solution gives insight into the origin of the gap at $p=0$, and how it closes for non-zero $p$.

At $J=0$, we have $\Sigma_{\psi_1 \psi_1} = \Sigma_{\psi_2 \psi_2} = 0$, and then the expressions in Section~\ref{sec:luttinger} constitute exact results for the frequency dependence of all Green's functions in terms of the chemical potentials and $R_{c \psi_1}$ and $R_{\psi_1 \psi_2}$.

Here we examine these expressions in the FL* phase, where we also set $R_{\psi_1 \psi_2} = 0$. Then the non-zero Green's functions are 
\bea
G_{cc} (z) &=& G_c^0 \left( z + \mu - \frac{R_{c \psi_1}^2}{z + \mu_{\psi_1}} \right) \nonumber \\
G_{\psi_1\psi_1} (z) &=& \frac{1}{z+\mu_{\psi_1}} + \frac{R_{c \psi_1}^2 G_{cc} (z)}{(z + \mu_{\psi_1})^2} \label{ex1}
\eea
where $G_c^0 (z)$ is given in (\ref{gc0}). It is useful to note the large $|z|$ limit
\beq
G_c^0 (|z| \rightarrow \infty) = \frac{1}{z} + \frac{t^2}{z^3} + \ldots\,, \label{ex2}
\eeq
which establishes that there are no poles in (\ref{ex1}) at $z=-\mu_{\psi_1}$.

Consider first the undoped insulating limit $p=0$, where we obtain a gapped phase. Particle-hole symmetry requires $\mu=0$ and $\mu_{\psi_1} = 0$. Examination of (\ref{ex1}) then shows that there is a gap in the spectrum for any $R_{c \psi_1}$ as shown in Fig.~\ref{fig:ex1}.

\begin{figure}
\includegraphics[width=3.22in]{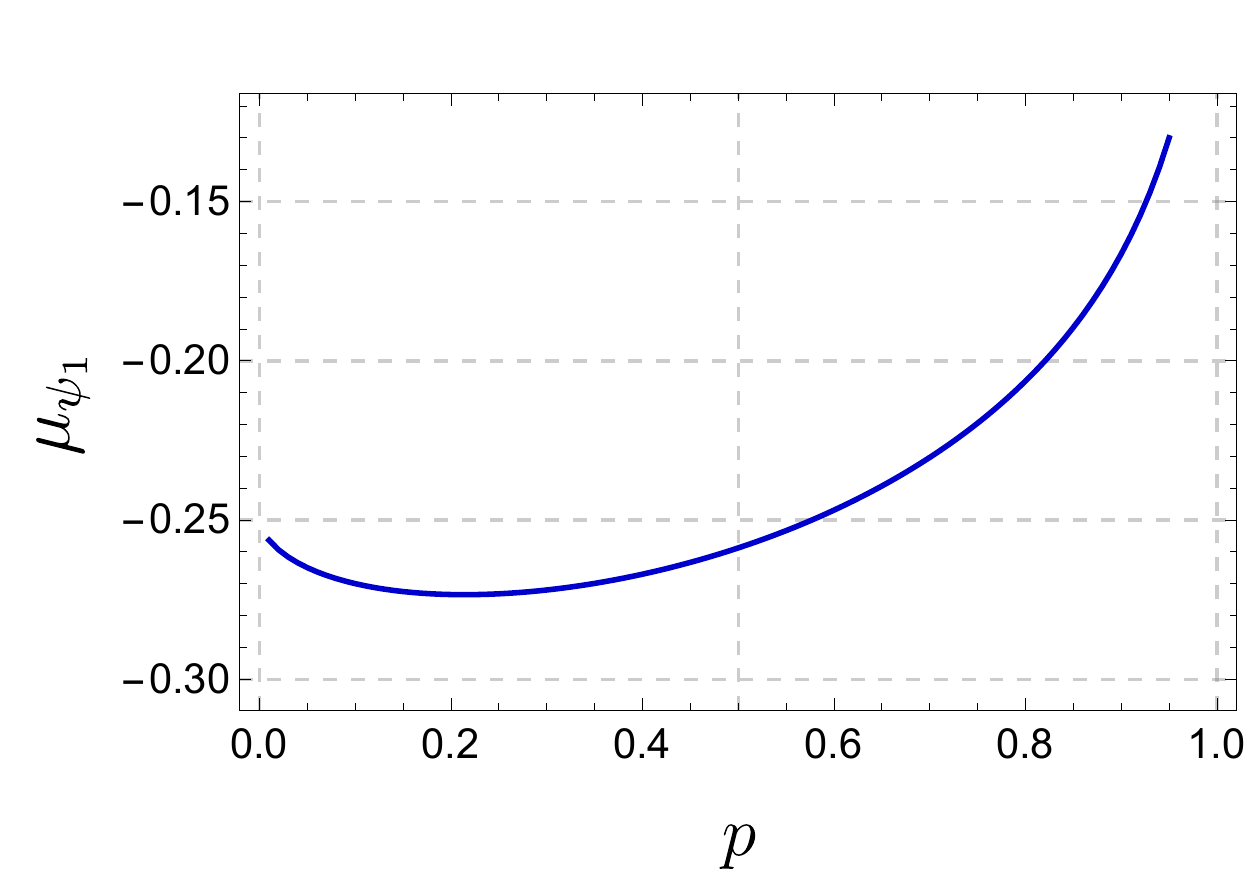}(a)\,\,
\includegraphics[width=3.12in]{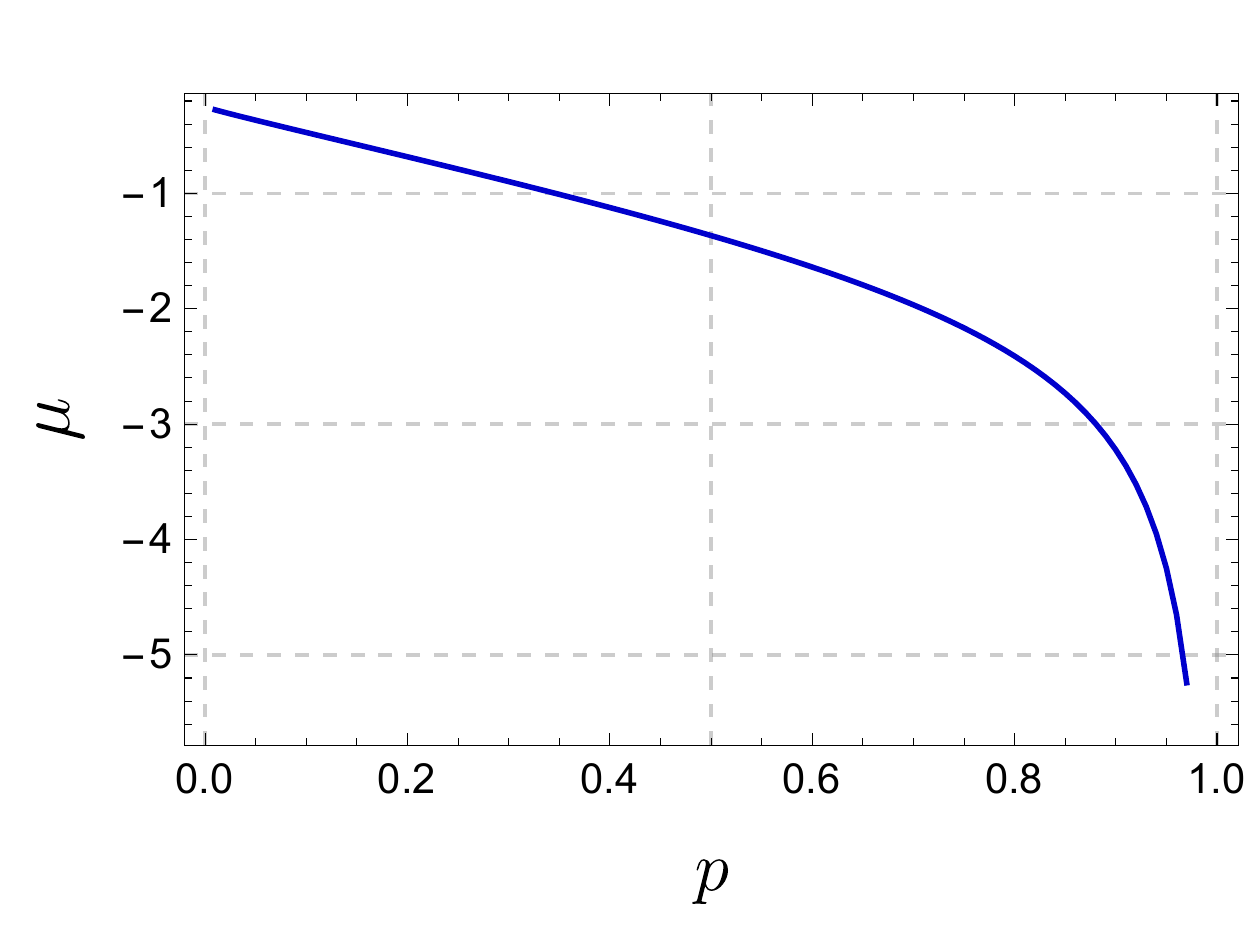}(b)
\caption{Numerically computed chemical potentials $\mu_{\psi_1}$ (a) and $\mu$ (b) as functions of doping $p$ for $J=0$, $t=1$, $R_{c \psi_1} = 0.75$.}
\label{fig:mu_p}
\end{figure}
\begin{figure}
\includegraphics[width=5in]{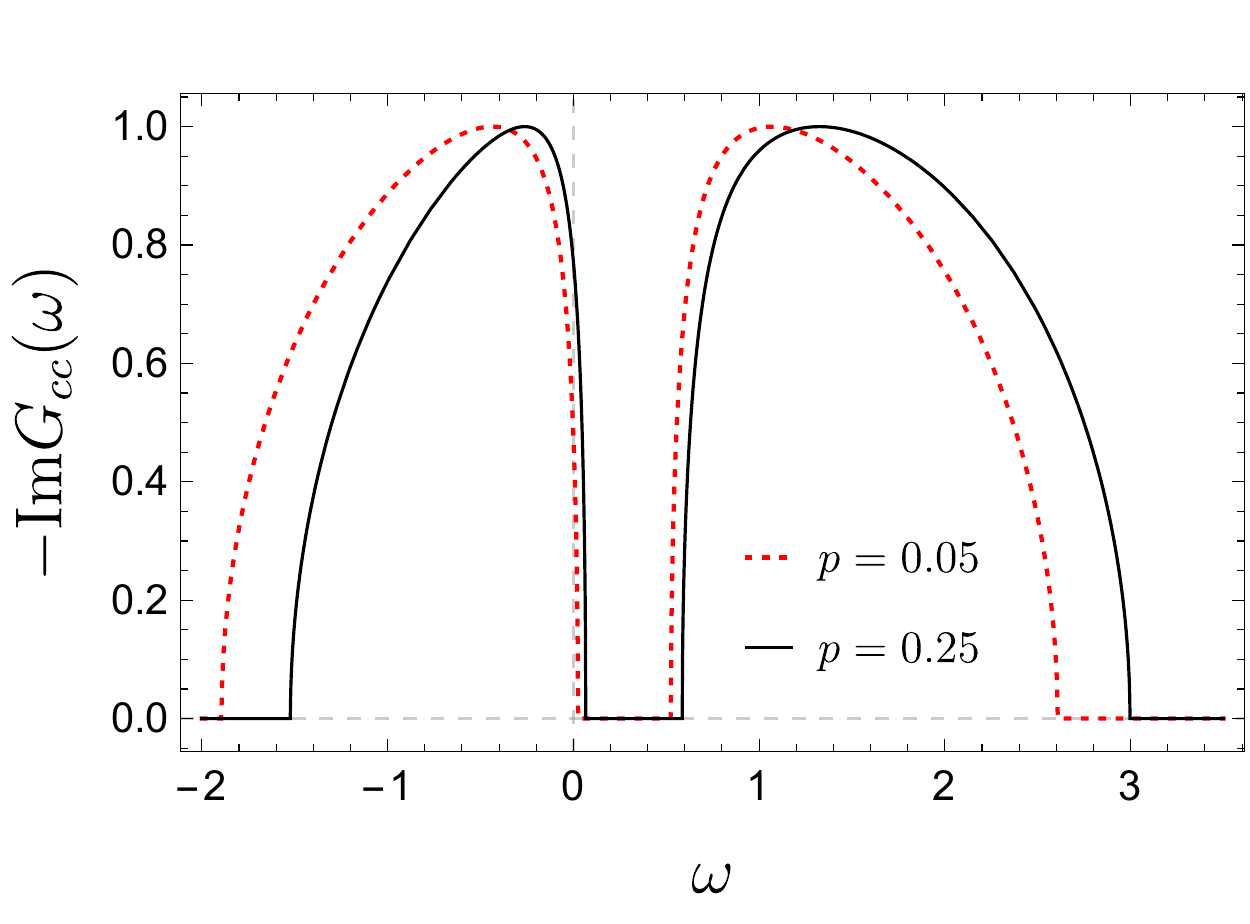}(a)\\
\includegraphics[width=5in]{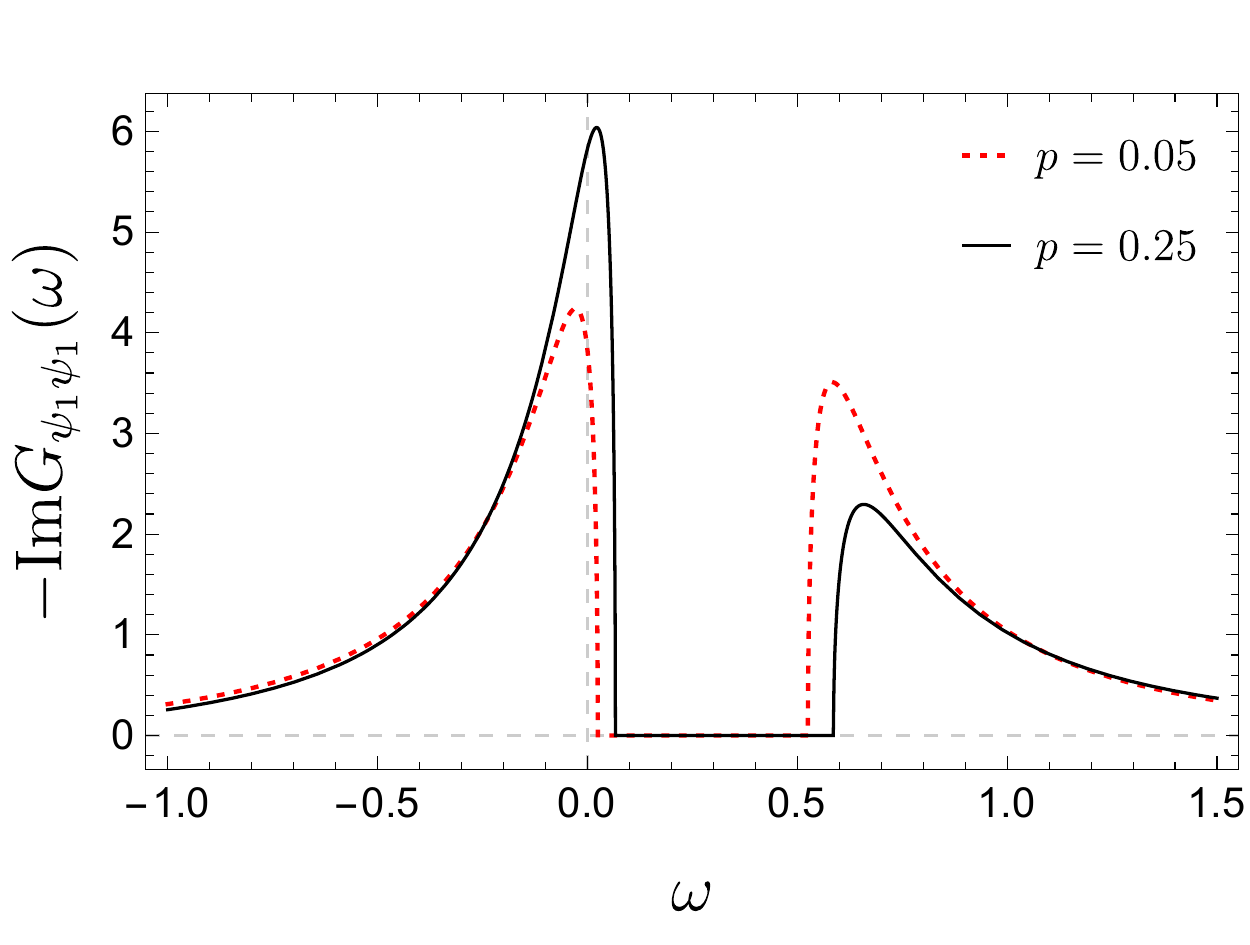}(b)
\caption{Electron (a) and fermionic (b) spectral densities at non-zero doping for $J=0$, $t=1$, $R_{c \psi_1} = 0.75$ and $p=0.05$, $\mu_{\psi_1}=-0.26$, $\mu=-0.37$ (red dotted line), $p=0.25$, $\mu_{\psi_1}=-0.27$, $\mu=-0.79$ (black solid line). }
\label{fig:rho_J0_p}
\end{figure}
Turning to non-zero $p$, with gapless metallic solutions. Now the Luttinger relation in (\ref{muval},\ref{EFval}) applies, and this relates $\mu$ to $p$
\beq\label{app:LR1}
\mu = E_F + \frac{R_{c \psi_1}^2}{\mu_{\psi_1}} \,,
\eeq
where $E_F \rightarrow 2t$ as $p \rightarrow 0$ as
\beq\label{app:LR2}
2 \int_{-2t}^{E_F} d \Omega D(\Omega) = 2-p \quad \Longrightarrow \quad \int_{E_F/(2t)}^1 dx \, \sqrt{1-x^2} = \frac{\pi p}{4} \,.
\eeq
At a given doping, the chemical potential $\mu_{\psi_1}$ is found numerically using the constraint \eqref{constraint1} and presented in Fig.~\ref{fig:mu_p}. 
The spectral densities change with doping as shown in Fig.~\ref{fig:rho_J0_p}. We check numerically the Luttinger relation \eqref{app:LR1} with the constraint \eqref{constraint2} and find that it works with the precision $10^{-4}$ which is comparable with the precision that we tune to find the chemical potential $\mu_{\psi_1}$. We also check the formula for the density of states of the electrons at the Fermi level \eqref{DoSEF} and obtain that it works with the precision $10^{-8}$.

\bibliography{pseudogap}

\end{document}